\documentclass{jfm}
\usepackage{graphicx}
\usepackage{epstopdf, epsfig}
\usepackage{hhline}
\usepackage{subfig}
\usepackage{caption}
\usepackage{natbib}
\usepackage{amsmath}
\usepackage{amssymb}
\usepackage{color}
\usepackage[dvipsnames]{xcolor}
\usepackage{bm}
\usepackage{hyperref}
\hypersetup{
    colorlinks = true,
    urlcolor   = blue,
    citecolor  = CadetBlue,
}

\shorttitle{Turbulent superstructures in thermal convection}
\shortauthor{S. Moller, T. Käufer, A. Pandey, J. Schumacher and C. Cierpka}

\title{Combined particle image velocimetry and thermometry of turbulent superstructures in thermal convection}

\author{Sebastian Moller\aff{1},
  Theo Käufer\aff{1}, Ambrish Pandey\aff{2}, J\"org Schumacher\aff{1,3}
 \and Christian Cierpka\aff{1}\corresp{\email{christian.cierpka@tu-ilmenau.de}}}

\affiliation{\aff{1}Institute of Thermodynamics and Fluid Mechanics, Technische Universität Ilmenau, D-98684 Ilmenau, Germany
\aff{2}Center for Space Science, New York University Abu Dhabi, Abu Dhabi 129188, UAE
\aff{3}Tandon School of Engineering, New York University, New York 11021, USA}

\begin{document}

\maketitle

\begin{abstract}
Turbulent superstructures in horizontally extended three-dimensional Rayleigh-B\'enard convection flows are investigated in controlled laboratory experiments in water at Prandtl number $Pr = 7$. A Rayleigh-B\'enard cell with square cross-section, aspect ratio $\Gamma = l/h = 25$, side length $l$ and height $h$ is used. Three different Rayleigh numbers in the range $10^5 < Ra < 10^6$ are considered. The cell is accessible optically, such that thermochromic liquid crystals can be seeded as tracer particles to monitor simultaneously temperature and velocity fields in a large section of the horizontal mid-plane for long time periods of up to 6 h, corresponding to approximately $10^4$ convective free-fall time units. The joint application of stereoscopic particle image velocimetry and thermometry opens the possibility to assess the local convective heat flux fields in the bulk of the convection cell and thus to analyse the characteristic large-scale transport patterns in the flow. A direct comparison with existing direct numerical simulation data in the same parameter range of $Pr, Ra$ and $\Gamma$ reveals the same superstructure patterns and global turbulent heat transfer scaling $Nu(Ra)$. Slight quantitative differences can be traced back to violations of the isothermal boundary condition at the extended water-cooled glass plate at the top. The characteristic scales of the patterns fall into the same size range, but are systematically larger. It is confirmed experimentally that the superstructure patterns are an important backbone of the heat transfer. The present experiments enable, furthermore, the study of the gradual evolution of the large-scale patterns in time, which is challenging in simulations of large-aspect-ratio turbulent convection.
\end{abstract}

\section{Introduction}
\label{sec:intro}
The occurrence of temperature differences and the resulting buoyancy forces induce complex fluid motions in many geo- and astrophysical settings and technical systems \citep{Kadanoff2001,Verma2018}. This is one reason why thermally driven fluid flows have been investigated comprehensively in the past decades in theory, simulation, and laboratory experiment. Of central interest is a better understanding of the close link between structure formation and the resulting turbulent heat transport properties. In this context, the Rayleigh-B\'{e}nard model has established as one of the most frequently considered basic configurations \citep{Bodenschatz2000,Ahlers2009,Chilla2012}. This simple convection model consists of a fluid layer between two parallel plates, which is uniformly heated from below and cooled from above, enclosed by adiabatic side walls. If the driving due to the vertical temperature gradient is large enough, turbulent Rayleigh-B\'{e}nard convection (RBC) develops in the layer that carries heat from the bottom to the top. Even though the model cannot represent the full complexity of most natural and technical systems, see e.g. \citet{Schumacher2020} for the case of convection in the solar interior, it turned out to be very useful for studies of fundamental turbulence characteristics in thermal convection, which also motivates our present study.

In this work, {\em turbulent superstructures of convection} -- a gradually evolving large-scale order in horizontally extended layers with a coherence length larger than the layer height~ --  are investigated in controlled laboratory experiments. The characteristic scale and dynamical evolution of superstructures is studied by a combination of stereoscopic particle image velocimetry and thermometry enabling a joint {\em experimental} determination of the local structure of the three velocity components $u_x$, $u_y$, and $u_z$ and the temperature $T$ as fields in a larger horizontal cross sectional plane. This opens the way to an experimental investigation of the spatial structure of the local convective heat flux, its fluctuations, and thus also the global heat transfer across the layer.  Our long-term experiments are conducted in water at a Prandtl number $\rm Pr=7$ in a cell with a square cross section and an aspect ratio $\Gamma=l/h=25$, with $l$ being the horizontal length and $h$ the height of the fluid layer. Furthermore, we compare the experimental results with existing direct numerical simulation data for the same parameter range which will be detailed later. Characteristic length scales and times are found to be in the same range as the numerical results. We also compare the statistics of the local convective heat flux, the velocity and temperature fields, and discuss differences that can be traced back to resolution effects in the experiments.   

The ``large-scale order'' in the form of superstructures in fully developed, horizontally extended convective turbulence has been analyzed with regard to the characteristic length scales for different Rayleigh and Prandtl numbers \citep{Hartlep2003,Parodi2004,Pandey2018,Pandey2021,Lenzi2021}, the associated dependence of the turbulent heat transfer on the aspect ratio of the convection layer \citep{BailonCuba2010}, and the effect of the aspect ratio on the characteristic pattern scale \citep{Hardenberg2008,Stevens2018}. In the studies of \citet{Emran2015} and \citet{Sakievich2016}, it has been demonstrated that the large-scale structures slowly rearrange over time. The characteristic lengths and time scales of the superstructures, as determined by \citet{Pandey2018} on the basis of numerical simulations in the Eulerian frame of reference, have also been confirmed in a complementary analysis of Lagrangian trajectory clusters \citep{Schneide2018,Vieweg2021a}. Deep learning algorithms have been applied to parametrize the turbulent convective heat flux in global models by a reduction to a planar dynamical network assessing the amount of heat transported by superstructures in \citet{Fonda2019}. Moreover, the coherence of the superstructures in the temperature and velocity field has recently been investigated in detail by \citet{Krug2020} and \citet{Blass2021}, while the interplay between small-scale turbulent fluctuations and the large-scale turbulent superstructures has been studied by \citet{Green2020}, \citet{Berghout2021}, and \citet{Valori2021}.

All studies of turbulent superstructures cited above are based on direct numerical simulations of the Boussinesq equations. The number of experimental investigations of turbulent convection in large aspect ratio cells with $\Gamma\gg 1$ is much smaller.  \citet{Fitzjarrald1976}  measured co-spectra of the temperature and velocity in air at a Prandtl number of $\rm Pr = 0.7$ with local sensor elements and determined a Rayleigh-number-dependence of the characteristic convection pattern scale. The large-scale structures in RBC have also been visualized in experiments with different working fluids at higher Prandtl and moderate Rayleigh numbers via the shadowgraph technique \citep{Busse1971,Busse1974,Busse1994} or photography \citep{Krishnamurti1981}. Fluctuation profiles of the velocity and temperature across the layer were investigated by \citet{Adrian1996}. Turbulent superstructure patterns have been obtained in convection experiments in air by \citet{Kaestner2018} and \citet{Cierpka2019} by time averaged velocity fields which were measured with particle image velocimetry, in short PIV \citep{Raffel2018}. 

Our experiment has been set up to perform simultaneous measurements of the temperature and velocity field over long time intervals in a large section of the horizontal midplane of a Rayleigh-B\'{e}nard cell. We therefore suspend thermochromic liquid crystals (TLCs) in the flow that change their color with respect to the local temperature \citep{Dabiri2008,Moller2021}. TLCs have also been used in RBC experiments by \citet{Zhou2007} to analyze the morphology of thermal plumes in a cylindrical cell with $\Gamma=1$. However, these particles serve in the current case not only as temperature sensors, but also as tracer particles for stereoscopic PIV measurements and thus enable the simultaneous velocity and temperature measurements, see also~\cite{Schmeling2014}. On the basis of these measurements, the strong influence of the turbulent superstructures on the local convective heat flux is demonstrated. In addition, their horizontal extent as well as their reorganization over longer time intervals is analyzed. In order to assess both the experimental and numerical investigations of turbulent superstructures in RBC, the results of the measurements are compared with those of direct numerical simulations (DNS) performed previously by \citet{Pandey2018} and \citet{Fonda2019}. 

The outline of the article is as follows. In \S~2, we present experimental details and give a short compact overview of the numerical simulations, which are used for comparison. \S~3 follows with a characterization of the turbulent superstructures. \S~4 presents the analysis of the local convective heat flux, which is followed by \S~5 and \S~6 on the characteristic length scales and long-term evolution of the large-scale patterns, respectively. The manuscript ends with a conclusion and an outlook in \S~7.

\section{Methods}
The turbulent Rayleigh-B\'{e}nard flow is determined by the geometry of the flow domain, the working fluid and the strength of the thermal driving. These dependencies are quantified by dimensionless numbers, the already mentioned aspect ratio $\Gamma=l/h$, the Prandtl number $\mathrm{Pr}=\nu/\kappa$ and the Rayleigh number $\mathrm{Ra}=\alpha g \Delta T h^3/(\nu \kappa)$. Here, $\Delta T$ is the temperature difference between the hot and cold plates at the bottom and top, respectively, $g$ is the acceleration due to gravity, and $\alpha$, $\nu$ and $\kappa$ represent the thermal expansion coefficient, the kinematic viscosity and the thermal diffusivity of the working fluid, respectively. Since the aspect ratio, the Prandtl number and the Rayleigh number determine the flow, these dimensionless numbers will be similar to those of the numerical simulations for comparison.

\subsection{Experimental investigation}
\label{sec:exp_inv}
\begin{figure}\centering
\includegraphics{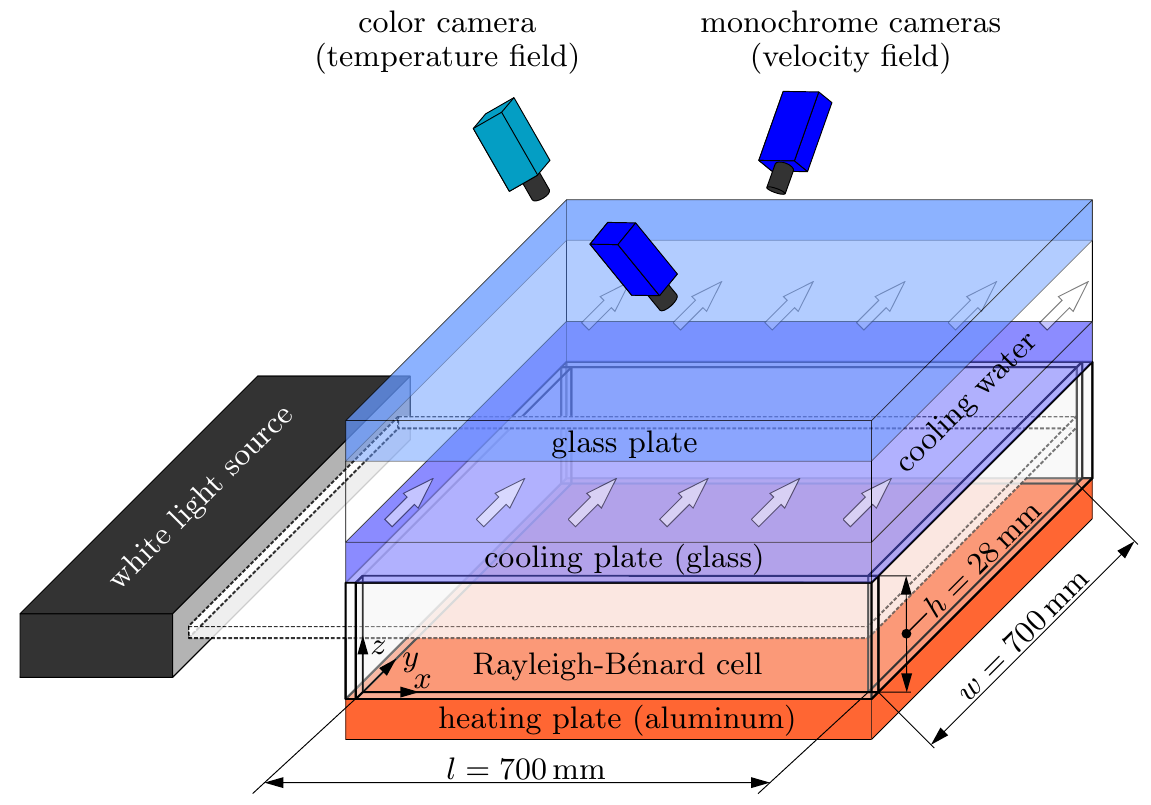}
\caption{Sketch of the experimental Rayleigh-B\'{e}nard convection facility.}
\label{fig:RBcell}
\end{figure}

For the experimental investigation of the turbulent superstructures in RBC a cuboid cell with dimensions of $l \times w \times h = 700\,\mathrm{mm} \times 700\,\mathrm{mm} \times 28\,\mathrm{mm}$, thus having an aspect ratio of $\Gamma=25$, has been built. The experimental setup is explained in detail in \citet{Moller2021}. A sketch of the experiment can be seen in figure~\ref{fig:RBcell}.

While the heating plate at the bottom of the cell is made of aluminum, the sidewall and the cooling plate at the top are made of glass. As can be seen in figure~\ref{fig:RBcell}, the temperature of the cooling plate is adjusted by an external cooling water circuit, which is covered with another glass plate at the top side. Since the working fluid in the cell is water for all the experiments in the present work, the entire flow domain is optically accessible, such that optical measuring techniques can be used to study the Rayleigh-B\'{e}nard flow. In our case, thermochromic liquid crystals (TLCs) are inserted into the flow as tracer particles to measure quantitatively the temperature distribution in the fluid. When illuminated with white light, these particles change their color appearance in dependence on the temperature and can therefore be used to measure the local temperature in the flow. Physical details of this mechanism can be found in \citet{Dabiri2008}. The TLCs are illuminated in the central horizontal bulk region through the transparent sidewall with a thin sheet of white light, which was collimated using Fresnel lenses with a minimal thickness between 3$\,$mm and 4$\,$mm over the entire field of view. More details on the design of the white light source and the light sheet optics can be found in \citet{Moller2020}. For additional background information, the reader is also referred to \citet{Schmeling2014} or \citet{Anders2020}. 

The color appearance of the TLCs is recorded from the top with a color camera through the glass plates and the cooling water. Based on the hue value $H$, which represents the pure color shade in the $HSV$-$\,$color space ($H$~-~Hue, $S$~-~Saturation, $V$~-~Value), the temperature is determined from local hue value calibration curves via linear interpolation \citep{Moller2021}. Here, TLCs of type R20C20W (LCR Hallcrest Ltd) are used. According to the nominal specifications, which are only achieved for the same direction of observation and illumination, the TLCs start to get red at $20\,^\circ$C and their color shade varies continuously across the visible wavelength spectrum until they appear in blue at $40\,^\circ$C. However, when analyzing the color of the TLCs dispersed in a fluid and illuminated with a thin light sheet, the observation angle between the color camera and the light sheet $\varphi_\mathrm{cc}$ must be varied towards an almost perpendicular arrangement. Since the applicable temperature range, in which the color shade changes from red to blue, shifts and decreases with increasing observation angles \citep{Moller2019,Moller2020}, the angle must be adapted carefully. In this case, the angle has been adjusted to $\varphi_\mathrm{cc}=65^\circ$ as a final tradeoff, which allows to measure the temperature in the temperature range $18.9\,^\circ\mathrm{C} \leq T \leq 20.8\,^\circ$C with a mean uncertainty of less than 0.1$\,$K \citep{Moller2021}.

Besides the color camera for the determination of the temperature field, two monochrome cameras in a stereoscopic arrangement are applied to record images of the TLCs in the horizontal mid-plane, such that not only the horizontal velocity components, but also the vertical velocity component can be measured by means of the stereoscopic particle image velocimetry \citep{Prasad2000,Raffel2018}. The main motivation for using the stereoscopic setup of the cameras is that the simultaneous measurement of the vertical velocity component and the temperature allows to estimate the local convective heat flux in RBC, as it will be shown in the following sections. In order to assess the effects of the Rayleigh number on the turbulent superstructures, three distinct temperature differences between the heating and cooling plates are prescribed, yielding the Rayleigh numbers which are listed in table~\ref{tab:param_exp} together with further experimental details. This experimental setup is appropriate to conduct long-term investigations of RBC -- a big advantage of experimental studies in comparison to direct numerical simulations, since it is not necessary to compute the current state of the flow with a high temporal and spatial resolution. As discussed in \citet{BailonCuba2010}, the numerical effort grows in addition with $\Gamma^2$. In the present case this allows to characterize the reorganization of the slowly evolving turbulent superstructures without determining the temperature and velocity field continuously, meaning that temporarily observing the flow with the cameras is sufficient to capture this gradual dynamics.

For each of the experiments, a strict procedure is followed. Once the temperatures of the heating and cooling plates reach the stationary state, the TLCs are inserted into the working fluid. As the flow in the cell is disturbed by the seeding process, we wait for 45$\,$min, so that the perturbation has decayed prior to the actual measurements. Afterwards, image series with a duration of 5$\,$min are recorded with a frequency of $f=5\,$Hz every $t_{\rm rec}=20\,$min. The last image series is recorded after six hours of continuous operation, resulting in 19 series with 1500 images for each Rayleigh number. In order to keep the computing time of the evaluation in limits, the temperature and velocity fields are determined for each fifth image. However, the images in between are necessary as well, because for the measurements at the two highest Rayleigh numbers the temporal delay of $\Delta t = 1\,$s between each fifth image is too large to properly compute the velocity field via PIV. On the contrary, the displacement of the TLCs on the cameras' images during the time gap of $\Delta t = 1\,$s has turned out to be well suited for the determination of the velocity field at the smallest Rayleigh number.
In addition, the temperature of both, the heating and the cooling plate was monitored by four Pt-100 temperature sensors in each plate during the whole experimental run to allow for a precise temperature control by the thermostats. The sensors that measure the heating plate temperature are mounted inside the aluminum heating plate close to the surface. Due to the heating plate's high thermal conductivity, this arrangement allows for a precise measurement of the heating plate temperature while not disturbing the flow. The temperature sensors at the cooling plate are glued on the glass surface that confines the fluid from the top. Thereby, the temperature of the cooling plate is measured directly at the solid-liquid interface.

The small temperature range of the TLCs and the intended large aspect ratio of the cell limit the range of accessible Rayleigh number ${\rm Ra}$ when turbulent convection under Boussinesq conditions is studied, see table \ref{tab:param_exp}.

\setlength{\tabcolsep}{1.9mm}
\renewcommand{\arraystretch}{1.3}
\begin{table}
\centering
\begin{tabular}{cccccccc}
Ra $/10^5$&  $T_\mathrm{h},T_\mathrm{c}$ in$\,^\circ$C  & $\tilde{x}_\mathrm{fov} \times \tilde{y}_\mathrm{fov}$ & $N_x \times N_y$ & $\Delta(\tilde{x},\tilde{y})$ & $t_\mathrm{f}\,$in s& $\tilde{\tau}_\mathrm{t}/10^3$ & $N_\mathrm{s}$ \\ \hline
$2.07$ & 19.78,$\,$19.08 & $16.1\times16.7$ &$191\times 198$ & 0.084 & 4.51& $5.39$ & $19\times299$ \\
$4.34$ & 20.22,$\,$18.76 & $16.1\times16.7$ & $191\times 198$ & 0.084 & 3.12& $7.79$ & $19\times300$\\
$7.34$ & 20.87,$\,$18.43 & $16.2\times16.6$ & $287\times 295$ & 0.084 & 2.40& $10.1$ & $19\times300$\\
\end{tabular}
\caption{Parameter sets for the measurements in the Rayeigh-B\'{e}nard cell with $\Gamma=25$ and water as the working fluid. The table lists the Rayleigh number Ra, the temperature of the heating plate $T_\mathrm{h}$ and cooling plate $T_\mathrm{c}$, the non-dimensionalized size of the intersection of the field of view (fov) of the three cameras $\tilde{x}_\mathrm{fov} \times \tilde{y}_\mathrm{fov} = x_\mathrm{fov}/h \times y_\mathrm{fov}/h$, the number of interrogation windows $N_x \times N_y$ with an overlap of 50\%, the resulting grid resolution $\Delta(\tilde{x},\tilde{y})$, the free-fall time unit $t_{\rm f}$, the non-dimensionalized total measurement time $\tilde{\tau}_\mathrm{t}~=~t_\mathrm{total}/t_\mathrm{f}$ after the initial transient period, and the number of the evaluated snapshots $N_\mathrm{s}$. It is noted that the Rayleigh numbers in the first column are always given as approximate values in the discussion of the results.}
\label{tab:param_exp}
\end{table}

\subsection{Direct numerical simulations for comparison}
\label{sec:num_inv}
In fluid dynamics, customarily non-dimensional or dimensionless equations are studied. Therefore, we simulate the following equations, which represent the conservation laws of mass, momentum and energy in dimensionless form where all fields are indicated by an additional tilde \citep{Chilla2012,Verma2018},
\begin{align}
\tilde\nabla \cdot \tilde{\bm u} & =0\,,
\label{eq:NS_mass}\\
\frac{\partial \tilde{\bm u}}{\partial \tilde t} + (\tilde{\bm u} \cdot \tilde\nabla)\,\tilde{\bm u} &= -\tilde{\nabla} \tilde {p} + \sqrt{ \frac{\mathrm{Pr}} {\mathrm{Ra}} } \, \tilde{\nabla}^2 \tilde{\bm u} + \tilde{T}\,{\bm e}_z\,,
\label{eq:NS_mom}\\
\frac{\partial \tilde T}{\partial \tilde t} + (\tilde{\bm u} \cdot \tilde\nabla)\,\tilde T &= \frac{1}{\sqrt{\mathrm{Ra}\mathrm{Pr}}} \tilde{\nabla}^2 \tilde T\,.
\label{eq:NS_en}
\end{align}
In these equations $\tilde{\bm u} = (\tilde{u}_x, \tilde{u}_y, \tilde{u}_z)$, $\tilde{p}$ and $\tilde{T}$ denote the dimensionless velocity, pressure and temperature fields, respectively. The equations are made dimensionless using the following scales: the height of the flow domain $h$ as the length scale, the temperature difference between the horizontal plates $\Delta T$ as the temperature scale, the free-fall velocity $u_f=\sqrt{\alpha g\Delta T h}$ as the velocity scale and the free-fall time $t_f=h/u_f$ as the time scale. For our incompressible flow, the pressure field is computed using the velocity field by solving a Poisson equation. Therefore, the characteristic unit of the pressure field is related to that of the velocity and reads $\rho_0 u_f^2$, with $\rho_0$ being the constant mass density of the fluid. All scales used for non-dimensionalization are given in table~\ref{tab:nondim}.

We integrate the equations (\ref{eq:NS_mass}) -- (\ref{eq:NS_en}) by applying a spectral element solver \textsc{Nek}5000 \citep{Fischer1997} in a cuboid box with square cross-section of size $25 h \times 25 h$. The velocity field satisfies the no-slip condition at all boundaries, while the temperature field satisfies the isothermal condition at the horizontal walls and the adiabatic (or thermally insulated) boundary condition at the vertical side walls. 
We decompose the flow domain into $N_e$ spectral elements which are again small cuboids. They form a non-uniform spectral element mesh with smaller element spacing towards all faces in order to resolve the strong variations of the velocity and temperature in the near-wall region correctly. The turbulence fields within each element are discretized using Lagrangian interpolation polynomials of order $N$ in each space direction using Gauss-Lobatto-Legendre collocation points. Thus, the flow domain consists of $N_eN^3$ mesh cells of non-uniform size, see also \citet{Scheel2013}.

With a view to the comparability of the numerical and experimental study of turbulent superstructures we use direct numerical simulation data of RBC in a flow domain with $\Gamma=25$ and $\mathrm{Pr}=7$. Furthermore, the Rayleigh number is adjusted to $\mathrm{Ra}=10^5$ and $\mathrm{Ra}=10^6$ in the simulations, because these Rayleigh numbers span the range of the experimental investigation.
The simulations are started from the motionless conduction state with random perturbations and we wait until the initial transients have decayed and the flow has reached a statistically steady state. After that, we continue our simulations for a total of $\tau_\mathrm{total} = $ 2670 free-fall time units for $\mathrm{Ra}=10^5$ and $\tau_\mathrm{total} =$ 3033 free-fall time units for $\mathrm{Ra}=10^6$. Finally, 268 and 560 equidistant snapshots were stored during this time interval for $\mathrm{Ra}=10^5$ and $\mathrm{Ra}=10^6$, respectively. Ensuring the adequacy of the spatial and temporal resolutions is important in order to faithfully study a turbulent flow using DNS. In a high-$\mathrm{Pr}$ convective flow, the smallest length scale in the temperature field is finer compared to that in the velocity field \citep{Silano2010, Pandey2014}.
Therefore, we ensure that the average size of the mesh cells remains smaller or comparable to the Batchelor length scale within each horizontal plane \citep{Scheel2013}. Furthermore, as the superstructures evolve on a much longer time scale compared to the free-fall time \citep{Pandey2018,Fonda2019}, we need to integrate the equations (\ref{eq:NS_mass}) -- (\ref{eq:NS_en}) for thousands of free-fall times, which poses additional restrictions on the numerical investigations of turbulent superstructures.
Further details about the DNS can be found in previous works by \cite{Pandey2018} and \cite{Fonda2019}.

\begin{table}
\centering
\begin{tabular}{lc}
Quantity & Characteristic unit\\ \hline
Length & $h$\\
Time & $t_f=h/u_f$\\
Velocity & $u_f=\sqrt{\alpha g\Delta T h}$\\
Temperature & $\Delta T$\\
Pressure & $\rho_0 u_f^2$\\
\end{tabular}
\caption{Non-dimensional characteristic units for all variables in the Rayleigh-B\'{e}nard convection flow.}
\label{tab:nondim}
\end{table}

\section{Characterization of turbulent superstructures}
\label{subsec:exposure}
\begin{figure}\centering\captionsetup[subfloat]{singlelinecheck=false,position={top},captionskip=-15pt,oneside,margin={1cm,0cm}}%
\subfloat[\label{fig:comp_exp_T_Ra7e5_inst}]{\includegraphics{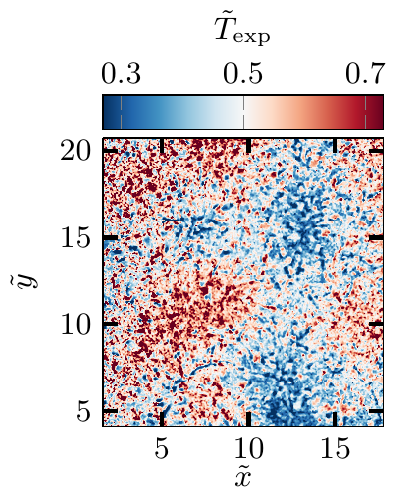}}\captionsetup[subfloat]{oneside,margin={0.2cm,0cm}}%
\hspace{-0.2cm}\subfloat[\label{fig:comp_exp_W_Ra7e5_inst}]{\includegraphics{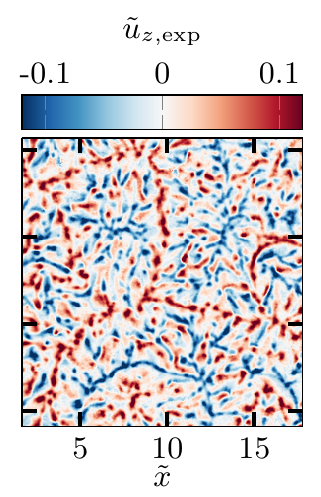}}%
\hspace{-0.2cm}\subfloat[\label{fig:comp_sim_T_Ra1e6_inst}]{\includegraphics{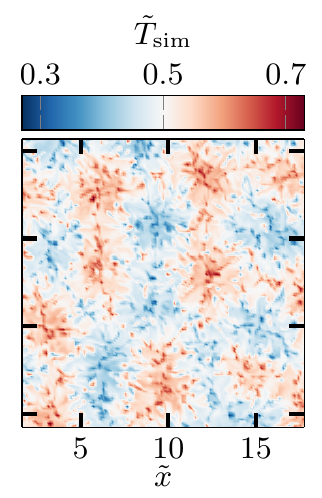}}%
\hspace{-0.2cm}\subfloat[\label{fig:comp_sim_W_Ra1e6_inst}]{\includegraphics{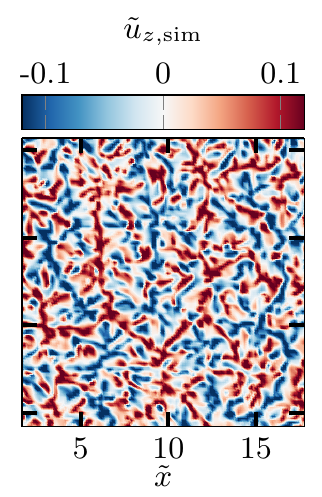}}\captionsetup[subfloat]{singlelinecheck=false,position={top},captionskip=-2pt,oneside,margin={1cm,0cm}}\vspace{-0.3cm}

\subfloat[\label{fig:comp_exp_T_Ra7e5_avg}]{\includegraphics{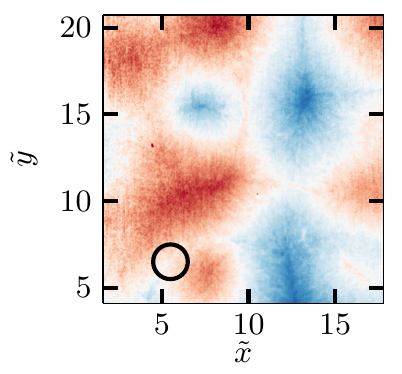}}\captionsetup[subfloat]{oneside,margin={0.2cm,0cm}}%
\hspace{-0.2cm}\subfloat[\label{fig:comp_exp_W_Ra7e5_avg}]{\includegraphics{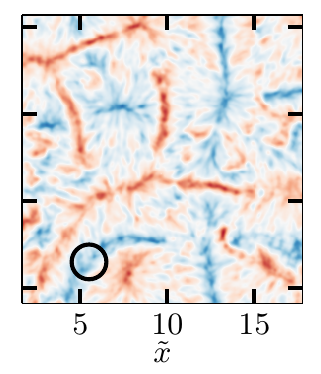}}%
\hspace{-0.2cm}\subfloat[\label{fig:comp_sim_T_Ra1e6_avg}]{\includegraphics{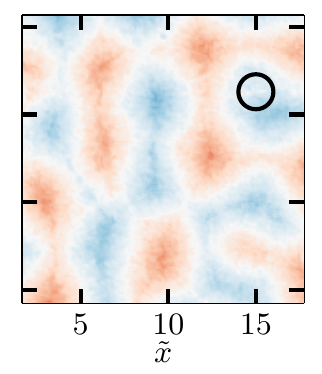}}%
\hspace{-0.2cm}\subfloat[\label{fig:comp_sim_W_Ra1e6_avg}]{\includegraphics{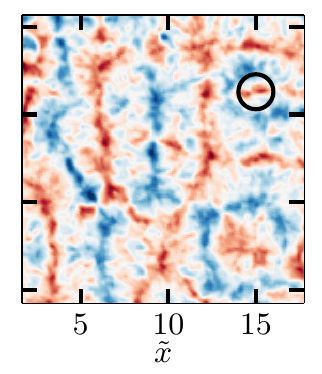}}%
\caption{Visualization of the instantaneous (a-d) and the corresponding time-averaged fields (e-h) of the temperature $\tilde T$ and of the vertical velocity component $\tilde{u}_z$ in the mid-plane, obtained from the laboratory experiments (exp) for the Rayleigh number ${\mathrm{Ra}=7 \times 10^5}$ (a,b,e,f) and from the direct numerical simulations (sim) for the Rayleigh number $\mathrm{Ra}= 10^6$ (c,d,g,h). The circles within the figures \ref{fig:comp_exp_T_Ra7e5_avg} - \ref{fig:comp_sim_W_Ra1e6_avg} indicate local events with a strongly pronounced vertical velocity component despite a small deviation from the average temperature as described in the text.}
\label{fig:comp_exp_sim}
\end{figure}

Exemplary instantaneous temperature and vertical velocity component fields are shown in figure~\ref{fig:comp_exp_sim}. The instantaneous fields in figures~\ref{fig:comp_exp_T_Ra7e5_inst} and \ref{fig:comp_exp_W_Ra7e5_inst} are obtained from measurements in the horizontal midplane at $\mathrm{Ra}=7 \times 10^5$, while those in figures~\ref{fig:comp_sim_T_Ra1e6_inst} and \ref{fig:comp_sim_W_Ra1e6_inst} result from the DNS at $\mathrm{Ra}=10^6$ at the same position. With respect to the measurement uncertainty the field of view of the three cameras applied for the measurements has been adjusted, yielding the intersection of the experimentally determined temperature and velocity fields in figure~\ref{fig:comp_exp_sim}. For a better comparability the numerical results are restricted to the same section of the horizontal midplane.

As it can be seen in figures~\ref{fig:comp_exp_W_Ra7e5_inst} and \ref{fig:comp_sim_W_Ra1e6_inst}, the Rayleigh-B\'{e}nard flow is characterized by strong spatial variations of $\tilde{u}_z$ with up- and downwellings due to rising and falling thermal plumes. The fingerprints of the flow can to some extent also be identified in the corresponding instantaneous temperature fields. However, one observes that larger areas with similar temperature ranges appear. A closer look reveals that these areas can also be detected in the instantaneous field of the vertical velocity component. This indicates that the flow does not only exhibit structures at smaller length scales, but also tends to develop patterns at somewhat larger scales. As the small-scale flow structures rearrange quickly and display stronger temporal fluctuations, they can be effectively removed in both fields by time-averaging. As a result, the large-scale patterns are revealed. The latter are denoted as turbulent superstructures and can be seen clearly in the time-averaged fields of the temperature and of the vertical velocity component in the lower row of figure \ref{fig:comp_exp_sim}. 

In the evaluation of the experimental data, the full recording time of 5$\,$min for each series of images has been taken as the time-averaging interval which translates into 125 free-fall units in the considered case with $\mathrm{Ra}=7 \times 10^5$. The numerical data have been evaluated according to \cite{Fonda2019}, in which averaging times of 207 free-fall units and 179 free-fall units are given for $\mathrm{Ra}=10^5$ and $\mathrm{Ra}=10^6$, respectively. Hence, the experimental data are averaged over a somewhat shorter time interval. However, the quantitative analysis on the basis of the time-averaged fields in \S~\ref{subsec:extent} and \S~\ref{subsec:reorganization} will not be affected by this slight difference, since the averaging time is sufficiently long to clearly reveal the turbulent superstructures, see also the supplementary information of \citet{Pandey2018}.

The comparison of the time-averaged temperature and velocity fields in figure~\ref{fig:comp_exp_sim} shows that the turbulent superstructures yield sharp contours in the vertical velocity component, while giving smoother contours of the temperature field. 
In addition, it can be anticipated without further analysis that the characteristic length scale in the temperature field is somewhat larger than in the velocity field. This behaviour was also found in the DNS by \citet{Pandey2018}. The effect can be explained by the fact that at the onset of convection, the temperature field and the vertical velocity field are perfectly synchronized (hot fluid dwells in upward direction whereas cold fluid sinks down) and thus show the same characteristic wavelengths. The synchronization decays with increasing Rayleigh number since the temperature field is not only advected by the vertical velocity component across the midplane, but also stirred by increasing velocity fluctuations in the horizontal directions. As a consequence the contours of the superstructures in the temperature field might appear broader given the fact that the thermal plumes are dispersed by the velocity fluctuations when rising or falling into the bulk of the layer. Locally this can lead to differences in the pattern scales of temperature and velocity fields. For example, this is visible around $(\tilde{x},\tilde{y})=(6,7)$ in the experiment and around $(\tilde{x},\tilde{y})=(15,16)$ in the simulation, where the temperature is close to $\tilde{T}=0.5$ despite a strong vertical fluid motion in upward or downward direction (see also circles in figure~\ref{fig:comp_exp_sim}). However, the turbulent superstructures in the temperature and velocity fields can be found in good qualitative agreement. The visual inspection of the results indicates that the experimentally determined turbulent superstructures seem to be somewhat larger, which will be detailed in \S~\ref{subsec:extent}.

\setlength{\tabcolsep}{3.5mm}
\renewcommand{\arraystretch}{1.15}
\begin{table}
\centering
\begin{tabular}{lccccc}
Ra & $10^5$ & $2 \times 10^5$ & $4 \times 10^5$ & $7 \times 10^5$ & $10^6$\\ \hline
Method & DNS & Experiment & Experiment & Experiment & DNS\\
$\sigma_T$ & 0.109 & 0.151 & 0.146 & 0.131 & 0.058\\
$\sigma_{T_\mathrm{co}}$ & 0.114 & - & - & - & 0.054\\
$U_\mathrm{rms}$ & 0.080 & 0.073 & 0.077 & 0.081 & 0.107\\
$U_{\mathrm{co},\,\mathrm{rms}}$ & 0.075 & - & - & - & 0.102\\
Re & 9.56 & 12.45 & 19.03 & 26.11 & 40.25\\
$\mathrm{Re} _{\mathrm{co}}$ & 8.96 & - & - & - & 38.36\\
Nu & 4.28 & 3.91 & 5.26 & 5.56 & 8.32\\
$\mathrm{Nu} _{\mathrm{co}}$ & 3.97 & - & - & - & 7.33\\
\rule{0pt}{2.3ex}$\langle \hat{\lambda}_T \rangle$ & 6.3 & 8.9 & 10.3 & 11.2 & 6.3\\
\rule{0pt}{2.3ex}$\langle \hat{\lambda}_{u_z} \rangle$ & 5.4 & 7.4 & 8.2 & 7.5 & 5.7\\
\end{tabular}
\caption{Results of the numerical and experimental investigations. The table lists the standard deviation of the temperature $\sigma_T$, the root-mean-square velocity $U_\mathrm{rms}$, the Reynolds number Re, the Nusselt number Nu according to its definition \eqref{Nu}, the mean characteristic wavelength of the turbulent superstructures determined from the temperature field $\langle \hat{\lambda}_T \rangle$ and from the field of the vertical velocity component $\langle \hat{\lambda}_{u_z} \rangle$. In case of the DNS, the quantities are determined from the horizontal mid-plane for a better comparability with the experimental data. In addition, the temperature and velocity data of the DNS are recomputed in small interrogation volumes around the horizontal mid-plane to estimate the effect of the smaller spatial resolution of the experimental data. The quantities resulting from the coarse-grained grid are denoted as $\sigma_{T_\mathrm{co}}$, $U_{\mathrm{co},\,\mathrm{rms}}$, $\mathrm{Re} _{\mathrm{co}}$ and $\mathrm{Nu} _{\mathrm{co}}$.}
\label{tab:evaluation_measurements}
\end{table}

Comparing the temperature fields in figure~\ref{fig:comp_exp_sim}, it can also be seen that the occurring temperatures span a wider range in the experiment. This is confirmed by the standard deviation of the temperature $\sigma_{T}$ in table~\ref{tab:evaluation_measurements}. We suspect that this is an effect of the turbulent superstructures on the temperature distribution at the heating and cooling plate in the experimental setup, as the impinging superstructures yield some slightly warmer and colder regions on both plates. In contrast, the range of vertical velocity amplitudes is larger in the simulations as obvious from figure~\ref{fig:comp_exp_sim}. A possible explanation for this observation is the higher spatial resolution of the DNS data. 
The in-plane resolution for the PIV measurements can be largely increased by advanced algorithms. For the current case we obtained an in-plane vector spacing of about $\Delta(\tilde{x},\tilde{y})$ = 0.084, which corresponds to 2.35 mm. In the vertical direction, the resolution is given by the light sheet thickness, amounts to 3 to 4 mm and cannot further be improved. It is this vertical resolution that limits a better resolution in all spatial directions.
Even though the experiment allows to resolve most of the small-scale structures, the peak velocities embedded in the fine structures are not fully captured due to the spatial averaging of the velocity in the particle image velocimetry measurements, such that the largest magnitudes are slightly smoothed out \citep{Kaehler2012}. 
However, the difference between the numerical and experimental results regarding the velocity range is smaller as seen in table \ref{tab:evaluation_measurements} and in figure~\ref{fig:Nu_Re_Ra}, where the root-mean square velocity $U_\mathrm{rms} = (\langle u_x^2+u_y^2+u_z^2 \rangle)^{1/2}$ and the resulting Reynolds number $\mathrm{Re}=(\mathrm{Ra}/\mathrm{Pr})^{1/2}\,U_\mathrm{rms}$ are also listed for each simulation and experiment. 

The root-mean-square velocities and Reynolds numbers of the DNS have also been computed on a coarser grid to investigate the effect of the smaller spatial resolution of the measurements. For this, the temperature and velocity data have been averaged in small interrogation volumes around the horizontal mid-plane. Hence, the numerical data have not only been averaged in the horizontal direction, but also to some extent in the vertical direction, thereby taking into account that the light sheet for the illumination of the measurement area has a certain thickness. The size of the interrogation volumes has been chosen such that the spatial resolution of the numerical data for $\mathrm{Ra}=10^5$ and $\mathrm{Ra}=10^6$ roughly matches that of the experimental data for $\mathrm{Ra}=2 \times 10^5$ and $\mathrm{Ra}=7 \times 10^5$, respectively. The resulting root-mean-square velocities $U_{\mathrm{co},\,\mathrm{rms}}$ and Reynolds numbers $\mathrm{Re} _{\mathrm{co}}$ are also given in table~\ref{tab:evaluation_measurements}. As expected, the magnitude of these quantities decreases due to the smaller spatial resolution. The numerical results are therefore fairly well in line with the experimental results.

\section{Analysis of the local convective heat flux}
\label{subsec:heat_flux}
The simultaneous measurement of the temperature and the vertical velocity component allows to determine the local convective heat flux in RBC.
Measurements of the local convective heat flux have been performed in the past, e.g., by tracking a mobile Lagrangian temperature sensor \citep{Gasteuil2007}.
\citet{Shang2004} combined laser-Doppler velocimetry (LDV) and temperature probe measurements to determine time series of the components of the local convective heat flux vector as a function of the cell radius. It was found that the most significant flux contribution stems from the large-scale circulation that rises or sinks along the sidewalls.
The novelty of the present study is that the local heat transfer is investigated as a field in the mid-plane in the Eulerian frame of reference. The Nusselt number ${\rm Nu}$ as the global measure of turbulent heat transfer is defined as 
\begin{equation}
{\rm Nu}(\tilde{z})=\sqrt{{\rm Ra Pr }}\,\langle \tilde{u}_z \tilde{T}(z)\rangle_{A,t} - \dfrac{\partial \langle \tilde{T}(z)\rangle_{A,t}}{\partial z} ={\rm const}\,.
\label{Nu}
\end{equation}
Recall that ${\rm Nu}(\tilde{z})$ takes the same value for any $0\le \tilde{z}\le  1$, even though the contributions to the heat transfer due to convective fluid motion (conv) and diffusion (diff) differ near the boundaries compared to those in the bulk, see e.g. \citet{Scheel2014}. At the horizontal mid-plane, the second term in eq. \eqref{Nu} is zero in the Boussinesq regime with its top-down symmetry. In dimensionless units, the two corresponding {\em local} currents are given by 
\begin{equation}
J_{\rm conv}(\tilde{\bm x}, \tilde t)=\sqrt{{\rm Ra Pr }}\, j_z(\tilde{\bm x}, \tilde t)=\sqrt{{\rm Ra Pr }}\, \tilde{u}_z(\tilde{\bm x}, \tilde t) \tilde{T}(\tilde{\bm x},\tilde t) 
\end{equation}
and
\begin{equation}
\quad J_{\rm diff}(\tilde{\bm x},\tilde t)=-\frac{\partial \tilde{T}(\tilde{\bm x},\tilde t)}{\partial \tilde z}\,.
\label{Nu1}
\end{equation}
Since $0\le \tilde T\le 1$, we take a symmetric form which is obtained by decomposing the original temperature field into the linear diffusive equilibrium profile and the rest, 
\begin{equation}
\tilde{T}(\tilde{\bm x},\tilde t) = 1-\tilde{z} +  \tilde\Theta(\tilde{\bm x},\tilde t)\,.
\label{Nu2}
\end{equation}
 Thus, we define eventually a {\em local convective} Nusselt number at $\tilde{z}=1/2$ by 
\begin{equation}
{\rm Nu}_\mathrm{loc}(\tilde x,\tilde y,\tilde t)=\sqrt{\mathrm{Ra}\mathrm{Pr}}\;\tilde{\Theta}(\tilde x,\tilde y,\tilde t)\,\tilde{u}_z(\tilde x,\tilde y, \tilde t)\,,
\label{Nu3}
\end{equation}
which is directly accessible as a field in the experiments. 
The uncertainty for the temperature measurements of 0.1 K results into an uncertainty for the individual values of the Nusselt number of about 14\% for Ra = $2 \times 10^5$, 7\% for Ra = $4 \times 10^5$ and 4\% for Ra = $7 \times 10^5$. However, the mean values (see table~\ref{tab:evaluation_measurements}) show a much lower uncertainty as $\sim 38000$ data points for the two lower Rayleigh numbers and $\sim 85000$ data points for the highest Rayleigh number are averaged.

\begin{figure}\centering\captionsetup[subfloat]{singlelinecheck=false,position={top},captionskip=-15pt,oneside,margin={0.9cm,0cm}}%
\subfloat[\label{fig:T_field_AV25_mid_Ens4_t_0_0_Ra2e5_nondim}]{\includegraphics{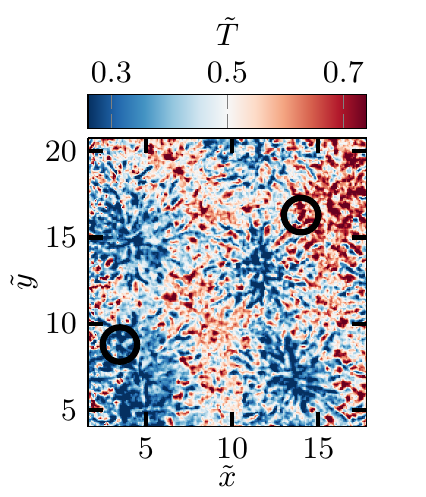}}\captionsetup[subfloat]{oneside,margin={0.6cm,0cm}}%
\hspace{-0.5cm}\subfloat[\label{fig:W_field_AV25_mid_Ens4_t_0_0_Ra2e5_nondim}]{\includegraphics{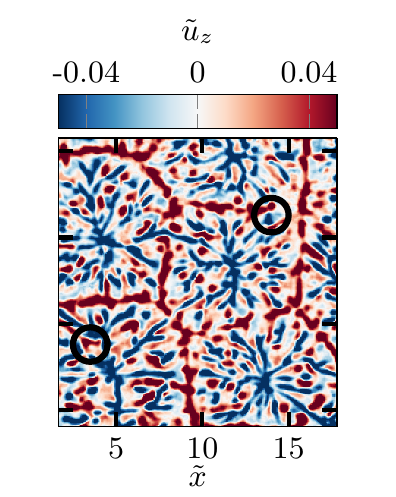}}%
\hspace{-0.5cm}\subfloat[\label{fig:Nu_field_AV25_mid_Ens4_t_0_0_Ra2e5_nondim}]{\includegraphics{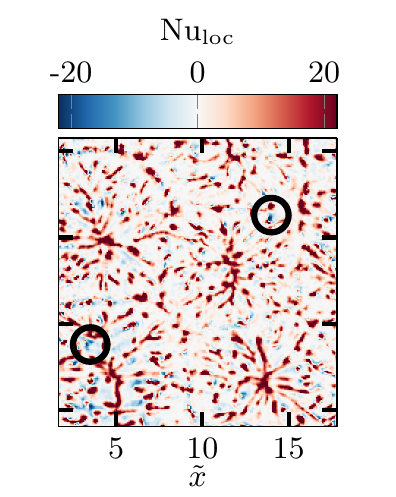}}\captionsetup[subfloat]{singlelinecheck=false,position={top},captionskip=-2pt,oneside,margin={0.9cm,0cm}}\vspace{-0.3cm}

\subfloat[\label{fig:T_field_AV25_mid_Ens4_t_0_298_Ra2e5_nondim}]{\includegraphics{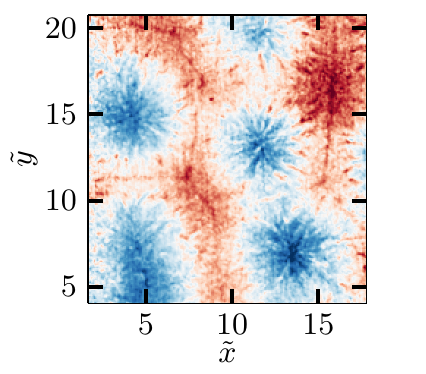}}\captionsetup[subfloat]{oneside,margin={0.6cm,0cm}}%
\hspace{-0.5cm}\subfloat[\label{fig:W_field_AV25_mid_Ens4_t_0_298_Ra2e5_nondim}]{\includegraphics{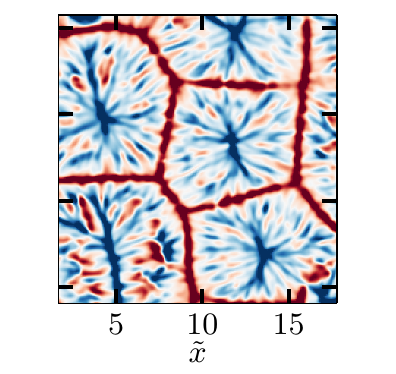}}%
\hspace{-0.5cm}\subfloat[\label{fig:Nu_field_AV25_mid_Ens4_t_0_298_Ra2e5_nondim}]{\includegraphics{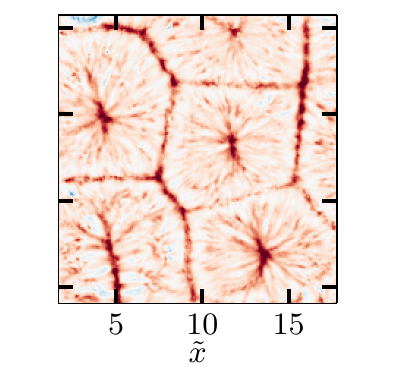}}\vspace{-0.2cm}
\caption{Exemplary instantaneous and the corresponding time-averaged fields of the temperature (a,d), the vertical velocity component (b,e) and the local Nusselt number (c,f) obtained from the measurement in the mid-plane at the Rayleigh number ${\mathrm{Ra} = 2 \times 10^5}$. The fields are computed as $\tilde{T} = (T-T_\mathrm{c})/(T_\mathrm{h}-T_\mathrm{c})$, $\tilde{u}_z = u_z/\sqrt{\mathstrut h g \alpha \Delta T}$ and $\mathrm{Nu}_\mathrm{loc} = \sqrt{\mathrm{Ra} \mathrm{Pr}} \,\tilde\Theta\,\tilde{u}_z$. The circles within the figures \ref{fig:T_field_AV25_mid_Ens4_t_0_0_Ra2e5_nondim} - \ref{fig:Nu_field_AV25_mid_Ens4_t_0_0_Ra2e5_nondim} indicate exemplary events with a negative local Nusselt number, i.e. the heat is transferred towards the heating plate as outlined in the text.}
\label{fig:T_W_Nu_diff_Ra}
\end{figure}

An exemplary instantaneous field of the local convective Nusselt number, which has been determined from the figures~\ref{fig:T_field_AV25_mid_Ens4_t_0_0_Ra2e5_nondim} and \ref{fig:W_field_AV25_mid_Ens4_t_0_0_Ra2e5_nondim}, is depicted in figure~\ref{fig:Nu_field_AV25_mid_Ens4_t_0_0_Ra2e5_nondim}. 
These results of the measurement at $\mathrm{Ra}=2 \times 10^5$ already indicate the effect of the superstructures on the local convective heat flux which appears even clearer in the corresponding time-averaged fields in the figures \ref{fig:T_field_AV25_mid_Ens4_t_0_298_Ra2e5_nondim}, \ref{fig:W_field_AV25_mid_Ens4_t_0_298_Ra2e5_nondim}, and \ref{fig:Nu_field_AV25_mid_Ens4_t_0_298_Ra2e5_nondim}. It can be seen that the maximum values of the local convective Nusselt number occur in the regions, where the time-averaged temperature reaches its extreme values, corresponding to updrafts and downdrafts of the turbulent superstructures. It is therefore confirmed that the superstructures strongly enhance the heat flux from the bottom to the top side of the Rayleigh-B\'{e}nard cell.
It can also be seen that multiple regions of strong heat transfer develop that are distributed homogeneously over the whole field of view. These regions are not constrained by the side walls which is different to Rayleigh-B\'{e}nard convection experiments in small aspect ratio cells \citep{Shang2004}.

\begin{figure}\centering\captionsetup[subfloat]{singlelinecheck=false,position={top},oneside,margin={1.3cm,0cm}}%
\hspace{1.35cm}\subfloat[\label{fig:PDFs_exp}]{\includegraphics{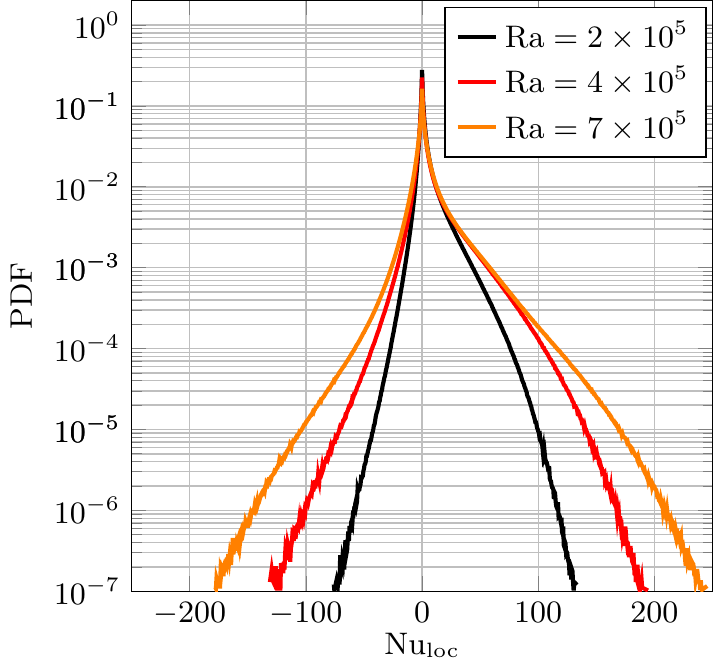}}\captionsetup[subfloat]{singlelinecheck=false,position={top},oneside,margin={0.2cm,0cm}}%
\subfloat[\label{fig:PDFs_sim}]{\includegraphics{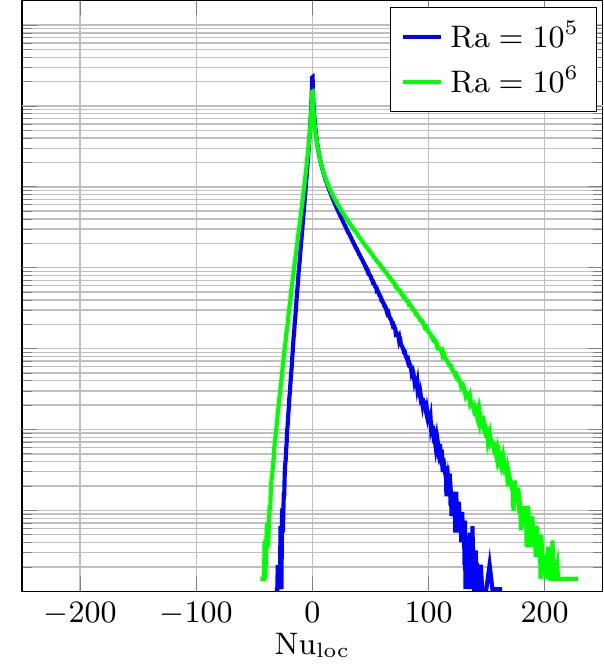}}\vspace{-0.2cm}
\caption{The probability density functions of the temporally resolved local convective Nusselt number resulting from the measurements in the mid-plane (a) and from the DNS~(b).}
\label{fig:PDFs_Nu}
\end{figure}

The quantitative analysis of the local heat flux is done by the probability density functions (PDFs) of the local convective Nusselt number as shown in figure~\ref{fig:PDFs_exp} for each of the three measurements and in figure~\ref{fig:PDFs_sim} for the two DNS. In this case, the PDFs incorporate all the temporally resolved values obtained from the measurements and the DNS at each Rayleigh number. As expected, the PDFs are more extended for positive values, since the total heat transfer increases due to convective motion. It can also be seen that the negative tails increase with the Rayleigh number, which shows that in turbulent thermal convection the upward motion of cold fluid and the downward motion of warm fluid become more probable. 

\begin{figure}\centering\captionsetup[subfloat]{singlelinecheck=false,position={top},oneside,margin={1.3cm,0cm}}%
\hspace{1.35cm}\subfloat[\label{fig:PDFs_j_rms_exp}]{\includegraphics{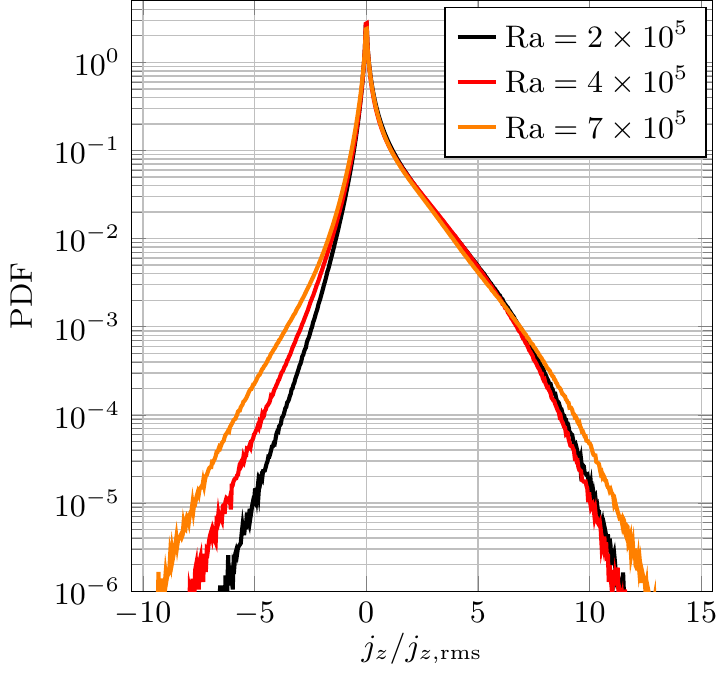}}\captionsetup[subfloat]{singlelinecheck=false,position={top},oneside,margin={0.2cm,0cm}}%
\subfloat[\label{fig:PDFs_j_rms_sim}]{\includegraphics{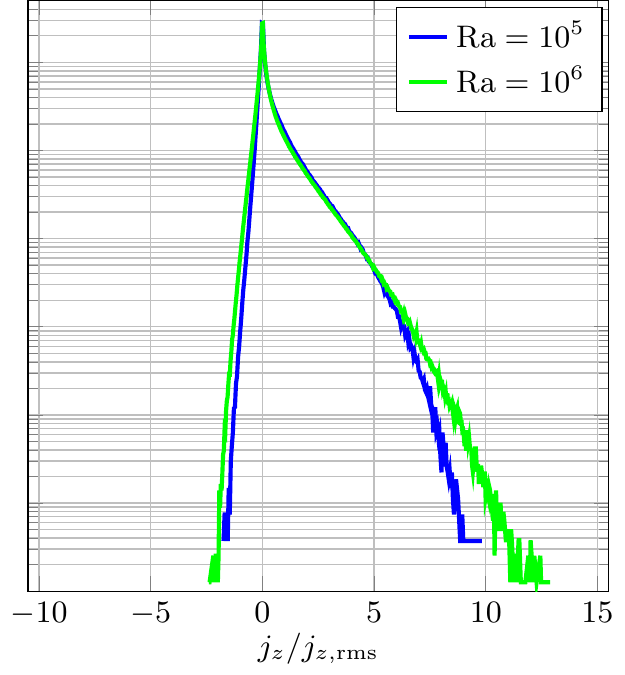}}\vspace{-0.2cm}
\caption{The probability density functions of the fluctuations of the vertical convective heat current component normalized by the corresponding root-mean-square value. These data result from the measurements in the mid plane (a) and from the DNS (b). Data are the same as in figure \ref{fig:PDFs_Nu}.}
\label{fig:PDFs_j_rms}
\end{figure}

The data also allows to evaluate the probability density functions of the normalized vertical convective heat current component $j_z/j_{z,{\rm rms}}$ as shown in figure~\ref{fig:PDFs_j_rms_exp} for each of the three measurements and in figure~\ref{fig:PDFs_j_rms_sim} for the two DNS. \citet{Shang2004} performed the same analysis and showed for a $\Gamma = 1$ cell that the PDFs fall on top of each other for a Rayleigh number range from Ra = $1.8\times 10^9$ to $7.6 \times 10^9$. The present PDFs of the normalized vertical convective heat current are supported on a similar interval (see their figure 9). The PDFs are also skewed to positive values since a mean heat flux goes from the bottom to the top. It can also be seen now that the positive tails on both panels collapse fairly well onto each other, both for experiments and simulations. However, the negative tails still differ even though being closer compared to figure \ref{fig:PDFs_Nu}. This effect is again attributed to the non-ideal boundary conditions that hinder the fast heat transport by conduction in the top plate. This effect becomes more dominant with increasing Rayleigh numbers as will be outlined in the following.

\begin{figure}\centering
\includegraphics[width=0.5\textwidth]{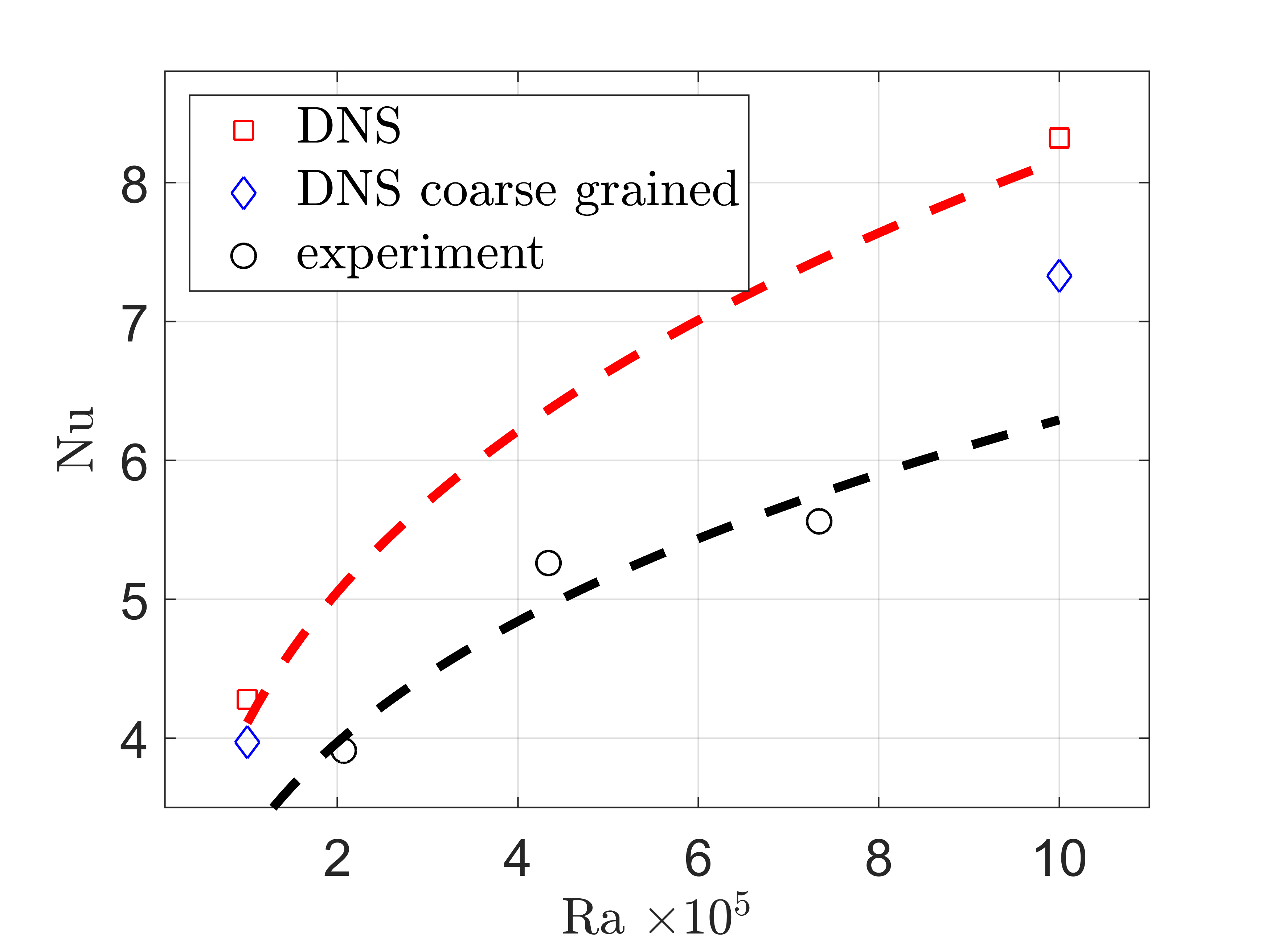}\vspace{-0.2cm}\includegraphics[width=0.5\textwidth]{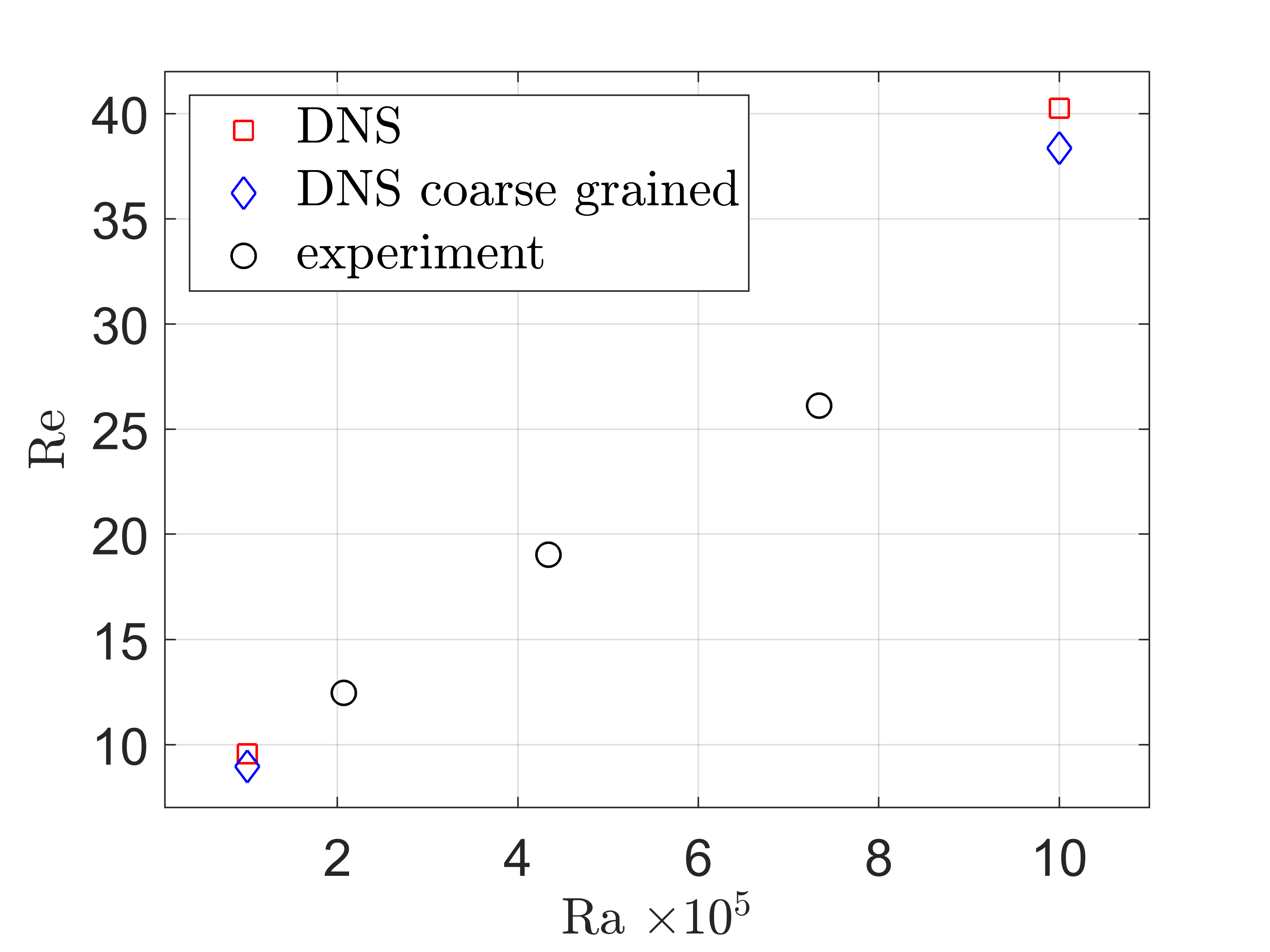}\vspace{-0.2cm}
\caption{Nusselt number vs. Rayleigh number (left) and Reynolds number vs. Rayleigh number (right). Values from table~\ref{tab:evaluation_measurements}, for the numerical results the coarse grained data are taken.}
\label{fig:Nu_Re_Ra}
\end{figure}

In order to further compare the experimental and numerical results regarding the heat flux, the spatial and temporal average of the local convective Nusselt number $\mathrm{Nu}_\mathrm{loc}$ 
are plotted in figure~\ref{fig:Nu_Re_Ra}. Even though the contribution of the diffusive heat flux cannot be obtained from the experimental data, the resulting average of the convective flux with respect to time and area is representative for the total heat flux, since the observation area in the experiment is large enough. By applying the power law fit of $\mathrm{Nu}=0.133\times\mathrm{Ra}^{0.298}$ (see red dashed line in figure~\ref{fig:Nu_Re_Ra}, left), which has been obtained from the DNS for three Rayleigh numbers of $\mathrm{Ra}=10^5$, $\mathrm{Ra}=10^6$ and $\mathrm{Ra}=10^7$ \citep{Fonda2019}, one gets $\mathrm{Nu}=5.11$, $\mathrm{Nu}=6.37$ and $\mathrm{Nu}=7.44$ for the three Rayleigh numbers of the experimental series. Hence, the global Nusselt number as determined from the experiments according to table \ref{tab:evaluation_measurements} is on average by about 25\% smaller than expected on the basis of the numerical results. Despite these deviations in magnitude, the scaling exponent of the global heat transfer power law can be considered as similar to the DNS. A power law fit of $\mathrm{Nu}=0.121\times \mathrm{Ra}^{0.286}$ results from the experimental data (see black dashed line in figure~\ref{fig:Nu_Re_Ra}, left).

An underestimated value in the experiment is possible, because the temperature and the velocity are measured with a lower spatial resolution and can also be considered as an average across the thickness of the light sheet used for the illumination of the tracer particles. The effect of these two aspects can be shown easily by means of the data of the numerical simulations. For this, the temperature and the vertical velocity component are averaged in small interrogation volumes around the horizontal midplane as it has been described in \S~\ref{subsec:exposure} for the estimation of the root-mean-square velocity $U_{\mathrm{co},\,\mathrm{rms}}$ and the Reynolds number $\mathrm{Re} _{\mathrm{co}}$. If the coarse-grained temperature and the vertical velocity fields are applied to compute the global Nusselt number $\mathrm{Nu} _{\mathrm{co}}$ from the DNS data, the values converge to the experimentally determined Nusselt number Nu according to the results in table~\ref{tab:evaluation_measurements}.

However, as the Nusselt numbers show, the convective heat flux in the experiment is still smaller in comparison to the simulations. This effect may be attributed to the boundary conditions. In the numerical simulations, perfect isothermal boundary conditions can be set, while temperature inhomogeneities at the boundaries cannot be fully impeded in an experiment. A dimensionless measure of the uniformity of the prescribed temperature at the boundaries is the Biot number~\citep{Verzicco2004,Schindler2021}, which is given by
\begin{equation}
{\rm Bi} = {\rm Nu}\left(\frac{k_{\rm f} h_{\rm w}}{k_{\rm w} h}\right),
\label{Bi}
\end{equation}
with $k_{\rm f}$, $h$ and $k_{\rm w}$, $h_{\rm w}$ being the heat conduction coefficient and the thickness of the fluid layer and of the wall material, respectively. The Biot number can also be interpreted as the ratio of the thermal resistance due to thermal conduction in the wall and the thermal resistance due to convective heat transfer at the wall. If ${\rm Bi}\ll 1$, then the limiting heat transfer mechanism is the convective heat transfer and not the conduction within the wall material, such that the isothermal boundary condition can be assumed. 

For the heating plate made of aluminum the Biot number is of the order of ${\rm Bi} \sim  10^{-3}$ and thus sufficiently small. However, the cooling plate is made of glass for the optical access. The glass plate has a thickness of $h_{\rm w} =8\,$mm and a thermal conductivity of $k_{\rm w} = 0.78\,$W/mK. Assuming $k_{\rm f} = 0.6\,$W/mK for the thermal conductivity of water at the mean fluid temperature of about $T=20\,^\circ$C, the Biot number increases with the Rayleigh numbers adjusted in the experiment from ${\rm Bi} = 0.86$ to ${\rm Bi} = 1.22$. Therefore, the thermal resistance due to conduction in the wall is not negligible in this case and limits the heat transfer. 
A transparent material with a much higher thermal conductivity would be sapphire. However, such a sapphire plate of 70 $\times$ 70 cm$^2$ is not available off the shelf and would have to be produced individually as a growing aluminum-oxide monocrystal. Therefore, we had to stick with the non-ideal boundary conditions.
Hot plumes from the bottom cannot transfer their thermal energy to the top plate in the same way as for the case of an isothermal boundary condition in the simulation. Due to continuity, the flow is forced to move downward with higher temperatures. This results in more events with a negative local convective Nusselt number as obvious from figure~\ref{fig:PDFs_exp}, where the correspondence between experiment and simulation is good for positive values of the local convective Nusselt Number, but differs for negative ones. The occurrence of events transferring heat towards the heating plate can clearly be seen in figure~\ref{fig:Nu_field_AV25_mid_Ens4_t_0_0_Ra2e5_nondim} in the form of blue regions, which correspond to hot fluid streaming  downward, for example at $(\tilde{x},\tilde{y}) \approx (14, 16)$, and cold fluid that moves upward,
for example at $(\tilde{x},\tilde{y}) \approx (4, 9)$, indicated by the circles in the figures \ref{fig:T_field_AV25_mid_Ens4_t_0_0_Ra2e5_nondim} - \ref{fig:Nu_field_AV25_mid_Ens4_t_0_0_Ra2e5_nondim}.

Due to the limitation of the convective heat transfer, which result from the material properties of the top plate, the local convective Nusselt number in the experiment is on average smaller than in the simulation, which is also visible by the larger positive tails for the DNS data in figure~\ref{fig:PDFs_Nu}. The difference for the mean value increases with the Rayleigh number, as the resistance for heat conduction in the top plate remains constant and the turbulent convective heat transfer increases and thus the Biot number increases as well. In the present case, the ratio of the Nusselt number obtained from the DNS and from the experiment is in the same order as the Biot number, i.e. ${\rm Nu}/{\rm Nu}_{\rm exp} \sim {\rm Bi}$. This effect would also explain the smaller velocities in the experiment, as the driving of the flow is hampered. Moreover, \cite{Vieweg2021} found that the variance of the temperature in the mid-plane is two orders of magnitude larger in the case of constant heat flux compared to the case of constant temperature at the horizontal boundaries~\footnote{This result is obtained by P.P. Vieweg through an additional analysis of simulation data published in \cite{Vieweg2021}}. This finding is in line with the larger standard deviation of the temperature $\sigma_{T}$ for the experimental data according to table~\ref{tab:evaluation_measurements}.

\section{Characteristic length scale of turbulent superstructures}
\label{subsec:extent}
The visual comparison of the temperature and velocity fields in figure~\ref{fig:comp_exp_sim} already indicates that turbulent superstructures in the experiments have a somewhat larger horizontal length scale than in the numerical simulations. For the quantitative estimation of the size of the superstructures in the experiment, the temperature field is considered in the following. As described in \citet{Pandey2018}, the size of the superstructures can be determined via Fourier spectral analysis on the basis of the field $\tilde\Theta$, which represents the temperature difference to the linear diffusive equilibrium profile at the height of the measurement plane. For simplicity, this field is denoted as the temperature field from here on. The characteristic horizontal length scale of the superstructures can be derived from the so-called power spectrum, which quantitatively represents the match between the temperature field and sinusoidal plane waves with different wavelengths and orientations.

\begin{figure}\centering\captionsetup[subfigure]{singlelinecheck=false,captionskip=-153pt,oneside,margin={0.85cm,0cm}}
\subfloat[\label{fig:procedure_Fourier_T_1}]{\includegraphics{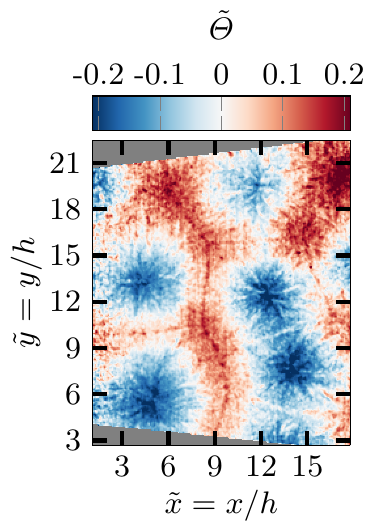}}%
\hfill\subfloat[\label{fig:procedure_Fourier_T_2}]{\includegraphics{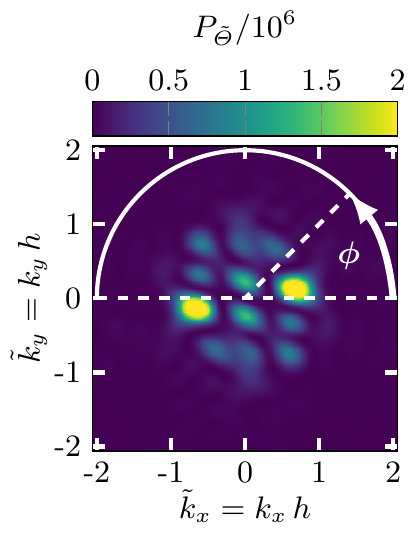}}\captionsetup[subfigure]{oneside,margin={1.2cm,0cm}}
\hfill\subfloat[\label{fig:procedure_Fourier_T_3}]{\includegraphics{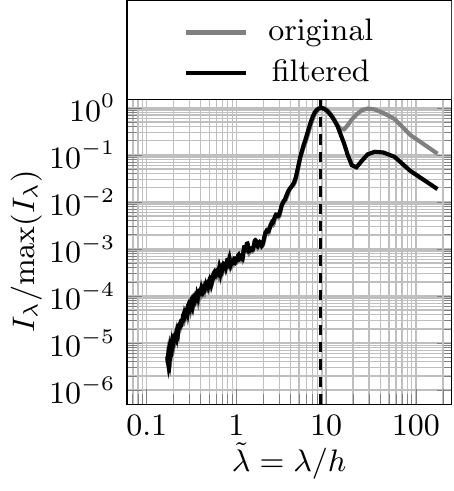}}\vspace{-0.2cm}
\caption{Procedure for the determination of the characteristic wavelength of the turbulent superstructures. (a) An exemplary time-window averaged temperature field $\tilde\Theta$, (b) the corresponding power spectrum $P_{\tilde\Theta}$, (c) the normalized wavelength intensities of the power spectrum determined from the original temperature field $\tilde\Theta$ and from the filtered temperature field $\tilde{\Theta}\big\rvert_{\,\forall\,\lvert\tilde{k}\rvert\,\geq\,0.4}$. The latter is shown in figure~\ref{fig:Shift_T_Ra2e5_select_1}.}
\label{fig:procedure_Fourier_T}
\end{figure}

A power spectrum of an exemplary time-averaged temperature field depicted in figure~\ref{fig:procedure_Fourier_T_1} can be seen in figure~\ref{fig:procedure_Fourier_T_2}. Here, the power spectrum $P_{\tilde{\Theta}}$ is arranged in the way, such that the dimensionless wavenumber $\tilde{k}=(\tilde{k}_x,\tilde{k}_y)$ increases from the center towards each of the edges. Hence, the entries of the power spectrum on the same circumference around the center represent features with the same wavenumber but a varying orientation. It will also be noted that the power spectrum is symmetric around the center, such that $P_{\tilde\Theta}\,(k_x,k_y)=P_{\tilde\Theta}\,(-k_x,-k_y)$. Due to the time-averaging, the turbulent superstructures represent the prominent pattern in the temperature field, which is also obvious from the two distinctive maxima of the power spectrum. These two symmetric maxima occur, since the superstructures are in this case preferentially arranged along a certain direction, namely along the line connecting the two maxima. However, not for all data records the superstructures exhibit a preferential orientation along a specific direction. In some cases, they are also represented by a ring-like structure of enhanced spectral power rather than two prominent maxima. Therefore, the azimuthal average of the power spectrum can be taken to determine the circumference with the largest average, in order to identify the characteristic wavelength of the superstructures. For a better visualization of the length scales the average intensities $I_\lambda$ normalized with the maximum are shown in figure~\ref{fig:procedure_Fourier_T_3} in dependence on the dimensionless wavelength $\tilde{\lambda}=\lambda/h$. The latter is related to the wavenumber via $\tilde{\lambda}=2\,\pi/\tilde{k}$. 

As the gray curve shows, there are two distinctive local maxima. While the peak at $\tilde{\lambda} \approx 8.7$ represents the superstructures, the peak at $\tilde{\lambda} \approx 29$ results from the experimental boundary conditions.
The second peak corresponds to the vertically aligned maxima close to the center in figure~\ref{fig:procedure_Fourier_T_2} and can be explained by the increasing temperature of the cooling water in its flow direction from the bottom to the top edge in figure~\ref{fig:procedure_Fourier_T_1}. 
A close look to this figure reveals that the measured temperature fields also exhibit the trend of larger temperatures along the cooling water flow in $y\,$-$\,$direction due to the slight increase of the cooling water temperature, which amounts to $\Delta T_{\rm{cw},y} \leq 0.15\,$K. In order to ensure that the characteristic wavelength of the superstructures is not determined based on the wrong maximum in any case, the large-scale trend is removed from the time-averaged temperature fields by computing the inverse Fourier Transform without considering the smallest wavenumbers close to the center of the power spectrum, meaning that the corresponding Fourier coefficients are set to zero. The limit of the wavenumber of $\tilde{k}=(\tilde{k}_x^2+\tilde{k}_y^2)^{(1/2)}=0.4$ has been chosen, such that all the wavelengths $\tilde{\lambda} > 2\pi/0.4=15.7$ are not taken into account. The latter span much wider horizontal dimensions compared to the turbulent superstructures. On the basis of the temperature field with the removed large-scale trend, which is depicted in figure~\ref{fig:Shift_T_Ra2e5_select_1}, the procedure for the determination of the average intensities $I_\lambda$ is then performed again. The resulting wavelength intensities with a clear peak at the characteristic wavelength of the superstructures can be also seen in figure~\ref{fig:procedure_Fourier_T_3}. This confirms the applicability of the filtering approach.

In the evaluation of the numerical data such a filtering is not required, because there are no deviations from the isothermal boundary condition yielding a large-scale trend of the temperature. Therefore, the power spectrum obtained from the time-averaged temperature field can directly be applied to derive the characteristic wavelength of the superstructures from the position of the maximum azimuthal average. The mean characteristic wavelength of the superstructures $\langle \hat{\lambda}_T \rangle$ obtained from the experimental and numerical investigations are given in table~\ref{tab:evaluation_measurements}. At each Rayleigh number the 19 time-averaged temperature fields resulting from the image series covering 5$\,$min have been considered to determine $\hat{\lambda}_T$ for the experiment, while the sliding time averages in the total runtime of the simulations have been used to calculate $\hat{\lambda}_T$ for the DNS. The total runtimes and the number of equidistantly stored sliding time averages are given in \cite{Fonda2019}.

For the estimation of the characteristic wavelength of the superstructures the time-averaged field of the vertical velocity component has been considered, too. In principle the resulting wavelengths $\langle \hat{\lambda}_{u_z} \rangle$ according to table~\ref{tab:evaluation_measurements} are obtained in the same way as for the temperature field, but the filtering approach of the experimentally determined velocity field has been changed slightly. It is obvious from figure \ref{fig:W_field_AV25_mid_Ens4_t_0_298_Ra2e5_nondim} that some regular variations of the vertical velocity component remain on small length scales after the time-averaging. Thus the filtering has been modified, such that variations of the velocity on either much larger or smaller length scales compared to the turbulent superstructures are removed. The wavelengths in the range $4.8 \leq \tilde{\lambda}_{u_z} \leq 15.7$ are not affected by the filtering. However, despite the additional filtering of small-scale variations of the velocity, the characteristic wavelengths determined from the time-averaged velocity field are somewhat smaller than those of the temperature field according to table \ref{tab:evaluation_measurements}. This is also the case for the fields obtained from the DNS, which might be caused due to the time averaging. 

The results in table~\ref{tab:evaluation_measurements} show that the characteristic wavelength of the superstructures $\langle \hat{\lambda}_T \rangle$ increases with the Rayleigh number according to the experiments, while the wavelength nearly remains constant according to the DNS. Moreover, the characteristic wavelengths determined from the DNS are typically one third smaller. Considering the boundary conditions of the experiment this is also reasonable. Due to a thick insulation around the sidewall of the cell and the adaption of the temperature in the lab to the mean temperature of the working fluid the adiabatic boundary condition at the sidewall is well satisfied, but deviations from the isothermal boundary condition at the horizontal plates cannot be fully impeded in an experimental setup. Despite extensive efforts, in the experimental setup at hand especially the top plate made of glass for the optical accessibility cannot compensate the effect of the impinging flow structures without a temporal delay. 

Based on the above discussion of the Biot number and on theoretical considerations~\citep{Hurle1967} it can be shown that the limited thermal conductivity of the plates enclosing the fluid layer strongly alters the flow in comparison to the classical RBC with isothermal boundary conditions. Essential is the ratio of the effective thermal conductivity of the working fluid $k_\mathrm{f}$ to the thermal conductivity of the solid plates $k_\mathrm{w}$. The two extreme cases $k_\mathrm{f}/k_\mathrm{w} \rightarrow 0$ and $k_\mathrm{f}/k_\mathrm{w} \rightarrow \infty$, which correspond to the \textit{Dirichlet boundary condition} with constant temperature at the plates and the \textit{Neumann boundary condition} with constant heat flux at the plates, have been compared in the study from \citet{Vieweg2021} on the basis of DNS for $\Gamma=60$. While the first case yields the turbulent superstructures with a horizontal extent larger than the height of the cell, the second case results in structures, which gradually grow and aggregate to one large structure spanning even the entire flow domain. None of both extreme cases is matched perfectly in any experiment, however, here the temperature at the bottom plate is almost uniform due to the high thermal conductivity of aluminum and $k_\mathrm{f}/k_\mathrm{w} \ll 1$. At the top plate which is made of glass, this ratio is $k_\mathrm{f}/k_\mathrm{w} \approx 1$ when taking into account that the effective thermal conductivity of the fluid increases due to the convective motion. Therefore, the increase of the horizontal characteristic scale of the superstructures in comparison to the DNS might be caused by non-ideal boundary conditions at the cooling plate which are a mixture of constant temperature and constant heat flux.

\section{Long-term evolution of the turbulent superstructures}
\label{subsec:reorganization}

\begin{figure}\centering\captionsetup[subfigure]{singlelinecheck=false,position={top},captionskip=-14pt,oneside,margin={0.85cm,0cm}}
\subfloat[\label{fig:Shift_T_Ra2e5_select_1}]{\includegraphics{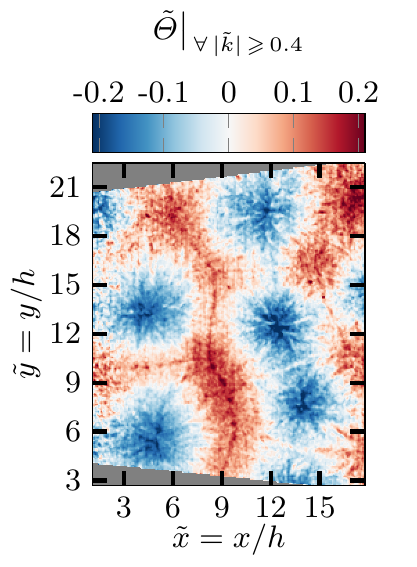}}\captionsetup[subfigure]{oneside,margin={0.45cm,0cm}}
\subfloat[\label{fig:Shift_T_Ra4e5_select_1}]{\includegraphics{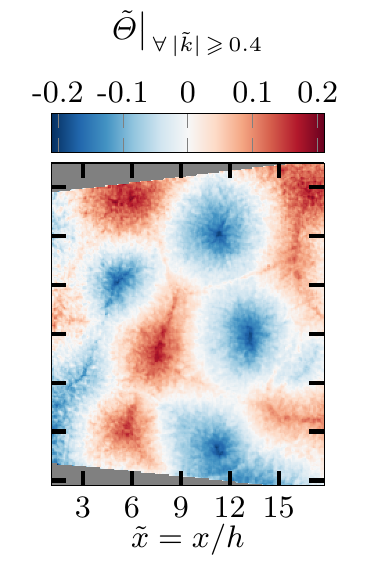}}
\subfloat[\label{fig:Shift_T_Ra7e5_select_1}]{\includegraphics{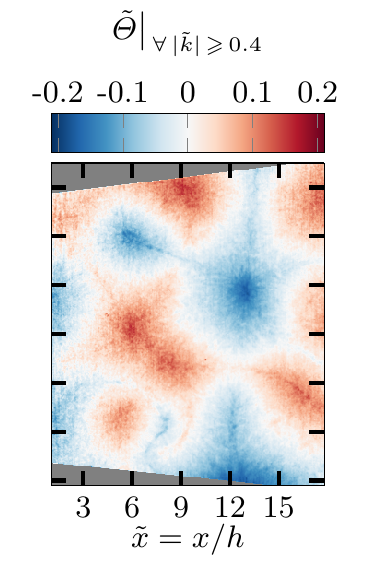}}\captionsetup[subfigure]{singlelinecheck=false,position={top},captionskip=0pt,oneside,margin={0.85cm,0cm}}\vspace{-0.3cm}

\subfloat[\label{fig:Shift_T_Ra2e5_select_2}]{\includegraphics{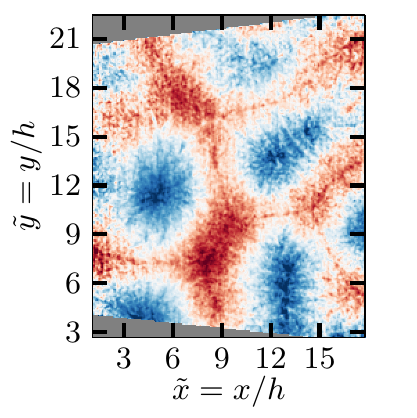}}\captionsetup[subfigure]{oneside,margin={0.45cm,0cm}}
\subfloat[\label{fig:Shift_T_Ra4e5_select_2}]{\includegraphics{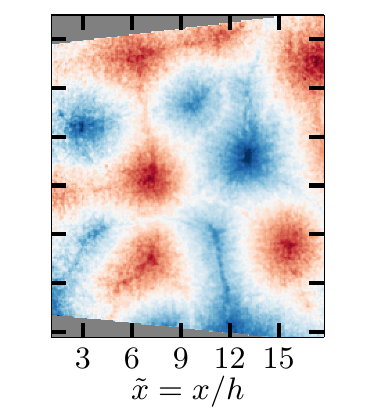}}
\subfloat[\label{fig:Shift_T_Ra7e5_select_2}]{\includegraphics{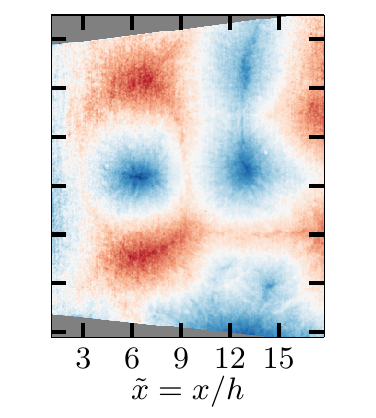}}
\caption{Exemplary demonstration of the reorganization of the turbulent superstructures by means of two time-averaged temperature fields in the mid plane at the Rayleigh number $\mathrm{Ra}=2 \times 10^5$ (a,d), $\mathrm{Ra}=4 \times 10^5$ (b,e) and $\mathrm{Ra}=7 \times 10^5$ (c,f).}
\label{fig:Reorganization_examples}
\end{figure}

Compared to the data of the DNS, which are often limited to total simulation times of several hundred free-fall units, the experimental data span approximately  $5\times 10^3$ up to $10^4 t_{\rm f}$ free-fall units for the different Rayleigh numbers. Hence, the ability to run the experiments over long time intervals opens the possibility to evaluate the reorganisation of the turbulent superstructures. The reorganization of the turbulent superstructures can be seen in figure~\ref{fig:Reorganization_examples}, which shows two different time-averaged temperature fields in the horizontal mid-plane for the different Rayleigh numbers adjusted in the experiments. The time difference between these two temperature fields is 200$\,$min, which corresponds to 2663 $t_{\rm f}$ at $\mathrm{Ra}=2 \times 10^5$, to 3846 $t_{\rm f}$ at $\mathrm{Ra}=4 \times 10^5$, and to 4988 $t_{\rm f}$ at $\mathrm{Ra}=7 \times 10^5$. The differences in the distribution of warm and cold regions in the temperature field demonstrate the relocation of the superstructures.

\begin{figure}\centering
\includegraphics[width=0.65\textwidth]{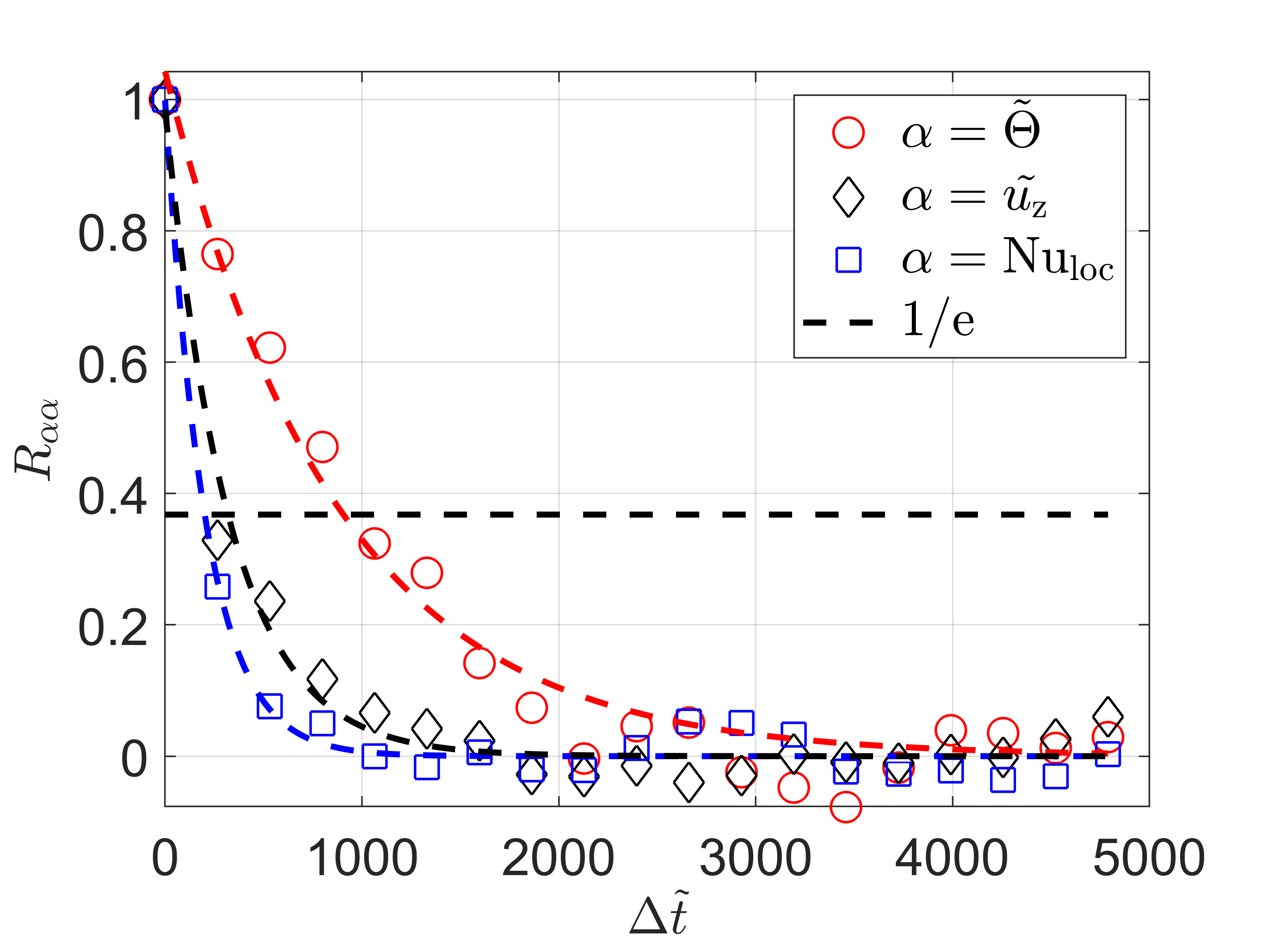}\vspace{-0.2cm}
\caption{Time correlation coefficient $R_{\alpha\alpha}$ for the experimental data obtained from the measurement at $\mathrm{Ra}=2 \times 10^5$. The typical time scales were estimated at a normalized correlation coefficient of $R_{\alpha\alpha}=1/e$, which is indicated with the horizontal dashed line.}
\label{fig:cc_2e5}
\end{figure}

On the basis of the numerical studies from \citet{Pandey2018} and \citet{Fonda2019} it can be estimated that the turbulent superstructures in water reorganize every 200 free-fall units on average, if the Rayleigh number is in the range $10^5 \leq \mathrm{Ra} \leq 10^6$. In order to assess the gradual long-term evolution of the superstructures, for each time-averaged field obtained from one of the 19 series of 1500 snapshots the central coefficient of the normalized two-dimensional cross-correlation is calculated according to
\begin{equation}
R_{\alpha\alpha}(\tilde{t}_0, \Delta\tilde t)=\frac{\langle (\bar{X}_{\alpha}(\tilde t_0)-\mu_{\bar{X}_{\alpha}(\tilde t_0)})(\bar{X}_{\alpha}(\Delta\tilde t)-\mu_{\bar{X}_{\alpha}(\Delta\tilde t)})\rangle}{\sigma_{\bar{X}_{\alpha}(\tilde t_0)}\sigma_{\bar{X}_{\alpha}(\Delta\tilde t)}} \quad \mbox{with}\enskip \Delta\tilde t=i\tilde{t}_{\rm rec}\,.
\end{equation}
Here, the running index $i \in [0,18]$ and $\tilde{t}_{\rm rec}=t_{\rm rec}/t_\mathrm{f}$ is the dimensionless time interval between each of the 19 image series with $t_{\rm rec}=20\,$min as given in subsection \ref{sec:exp_inv}.  The Greek symbol $\alpha$ represents the quantities $\alpha=\{\tilde T, \tilde u_z, {\rm Nu_{loc}}\}$, $\bar X_{\alpha}$ the respective time-averaged, dimensionless fields, $\mu_{\bar{X}_{\alpha}}$ and $\sigma_{\bar{X}_{\alpha}}$ the corresponding mean value and the standard deviation, respectively. Furthermore, $\langle\ \cdot\ \rangle$ denotes the average. This analysis shows that for all Rayleigh numbers the temporal variation of the time-averaged temperature fields is not as pronounced as the variation of the time-averaged fields of the vertical velocity component and the local Nusselt number. If two successive time-averaged temperature fields are considered, respectively, the correlation coefficient remains on a high level of $R_{\tilde{\Theta}\tilde{\Theta}} \approx \{0.5, 0.7, 0.8\}$ for $\rm Ra = \{2 \times 10^5, 4 \times 10^5, 7 \times 10^5\}$. This behaviour can again be attributed to the boundary conditions, as the deviation from the isothermal boundary condition at the top wall becomes stronger with increasing Rayleigh number. If the temperature value at the top plate is at a certain point larger than in the neighbourhood due to an impinging superstructure, this causes a stabilization of the superstructure at that position, such that a random reorganisation of the structures will take longer or is inhibited. In order to analyze the gradual reorganisation, the time-averaged fields of the temperature, vertical velocity component and local Nusselt number for $\mathrm{Ra}=2 \times 10^5$ are evaluated, as the dimensionless temporal delay between each of the fields is the smallest for this Rayleigh number and amounts to about $\tilde{t}_{\rm rec}=266$. The correlation coefficients have been normalized for this purpose and are shown in figure~\ref{fig:cc_2e5}.

Based on the normalized correlation coefficients, the typical time scales for the rearrangement of the superstructures are estimated to be $\tau_{\tilde T} \approx 920$, $\tau_{\tilde u_z} \approx 320$ and $\tau_{\rm{{Nu_{\rm{loc}}}}} \approx 200$ for the fields of the temperature, the vertical velocity component and the local Nusselt number, respectively. This confirms the aforementioned finding that in the present case the temperature patterns persist much longer in comparison to the characteristic patterns of the velocity field and of the corresponding local Nusselt number field. In order to further investigate the reorganisation of the turbulent superstructures and evaluate their role in heat transfer, a spatial and temporal median filter with a kernel size of $n_{\tilde{x}} \times n_{\tilde{y}} \times n_{\tilde{t}} = 9 \times 9 \times 3$ was applied to the total data. In the smoothed fields of the temperature, the vertical velocity component and the local Nusselt number isosurfaces can be extracted that display the reorganisation in time effectively.

\begin{figure}\centering
\includegraphics[width=0.48\textwidth]{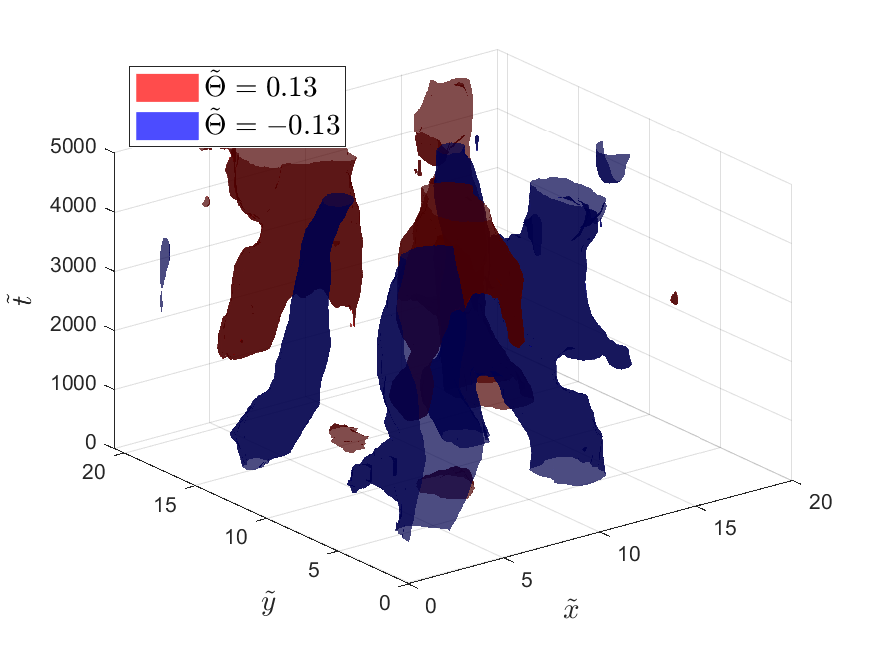}\vspace{0.1cm}
\includegraphics[width=0.48\textwidth]{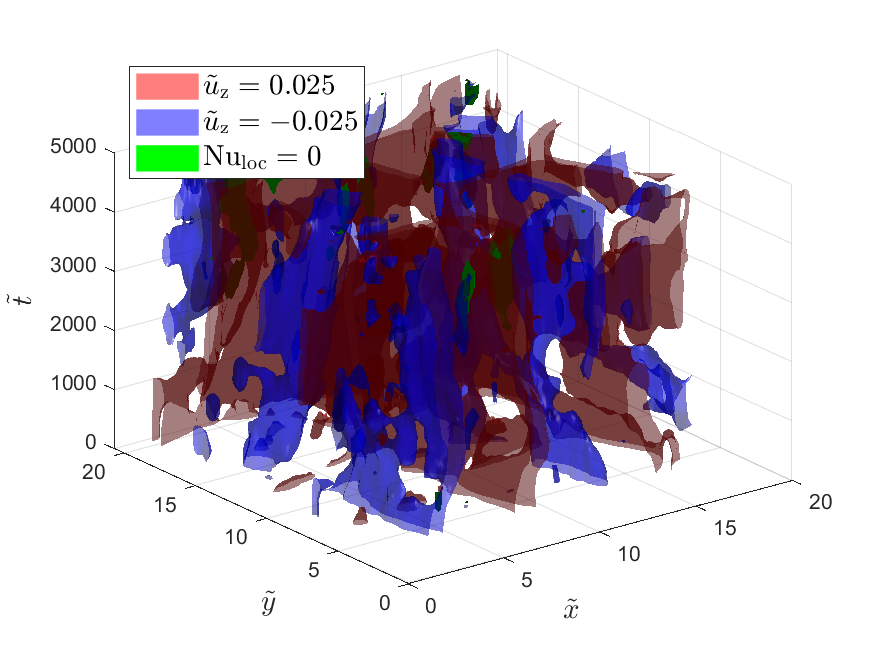}
\includegraphics[width=0.45\textwidth]{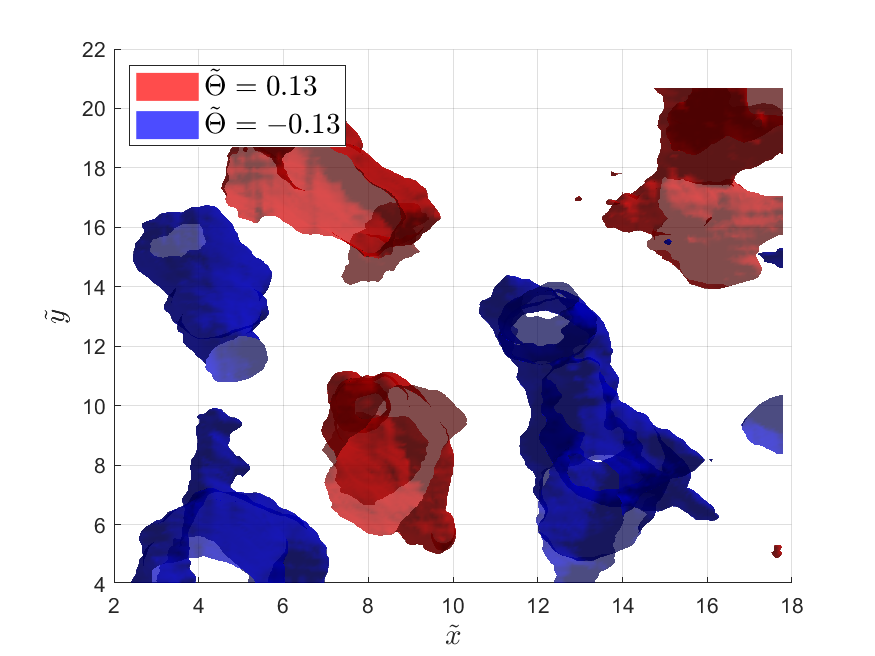}
\vspace{0.1cm}
\includegraphics[width=0.45\textwidth]{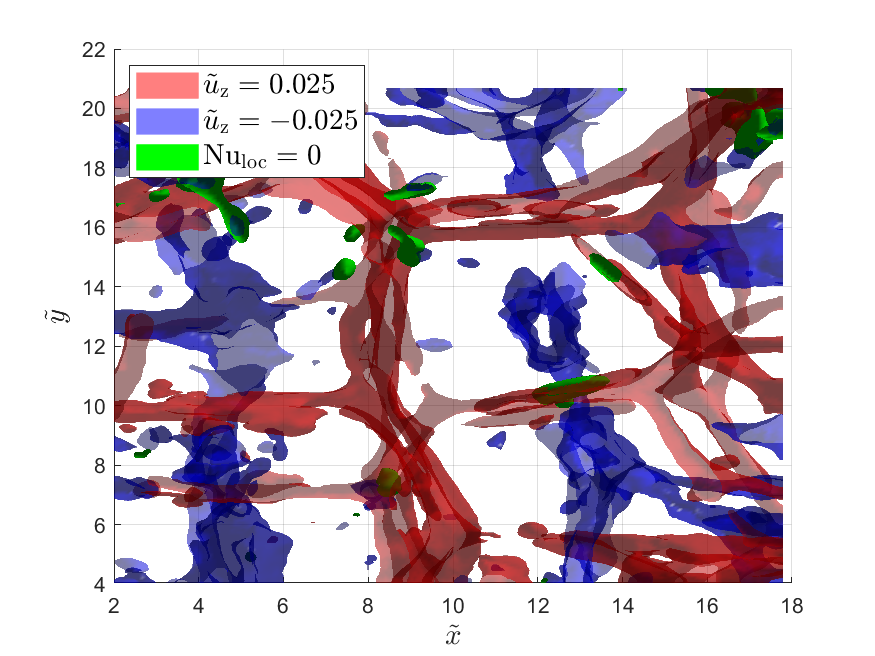}
\caption{Isosurfaces of $\tilde{\Theta}$ (left) as well as of $\tilde{u}_{\rm{z}}$ and $\rm{Nu}_{\rm{loc}}$ (right) for the experimental data obtained from the measurement at $\mathrm{Ra}=2 \times 10^5$. The lower row shows the isosurfaces from the top view.}
\label{fig:Isosurfaces}
\end{figure}

In figure~\ref{fig:Isosurfaces}, space-time isosurfaces of the time-averaged temperature field are shown for $\bar X_\Theta = \pm 0.13$ at the top left and of the time-averaged vertical velocity component field for $\bar X_{u_{z}} = \pm 0.25$ at the top right. Both fields are replotted in the lower row of the figure as a view from the top. In addition, isosurfaces of $\bar X_{\rm Nu_{loc}} = 0$ are superposed in green colour within the panels showing the velocity isosurfaces. It is seen that the temperature field contours change gradually due to a certain reorganization of the superstructures. The isosurfaces partly bifurcate or end in the course of the time evolution. Furthermore, the isosurfaces of the vertical velocity component on the right side of the figure vary much stronger in space and time as it can be expected on the basis of the correlation coefficients in figure~\ref{fig:cc_2e5}. In the view from the top, it can now be seen clearly that cells of strongly downwelling fluid in the middle (blue) and upwelling fluid in a kind of rim (red) do form. Even though these patterns remain relatively stable, especially for the case of the upwelling fluid a variation of the orientation of the ridges becomes apparent. These patterns do not necessarily coincide with the region of a larger temperature magnitude, which explains the decreased average of the local Nusselt number and the larger portion of negative local Nusselt numbers in the experiments, as obvious from table \ref{tab:evaluation_measurements} and figure~\ref{fig:PDFs_Nu}. It is interesting to note, that regions with negative local Nusselt numbers also remain for a certain time $\tilde t \sim {\cal O}(10^3)$.

\begin{figure}\centering
\includegraphics[width=0.48\textwidth]{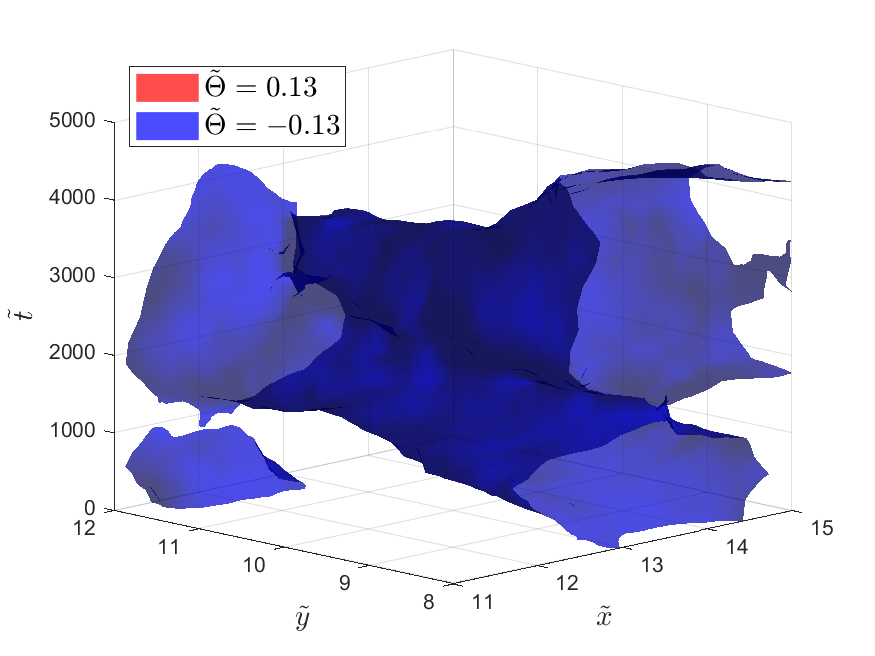}\vspace{0.1cm}
\includegraphics[width=0.48\textwidth]{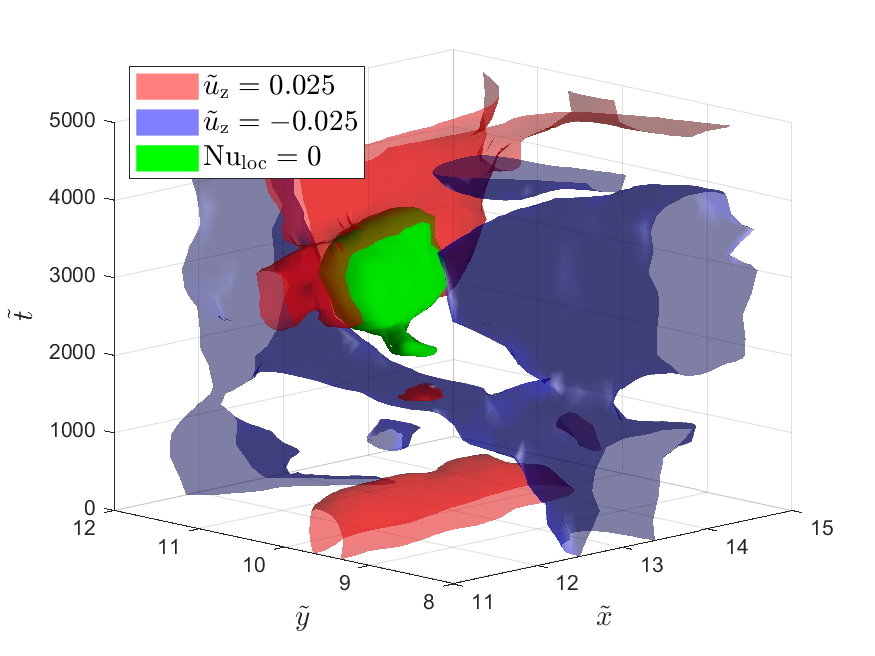}
\includegraphics[width=0.48\textwidth]{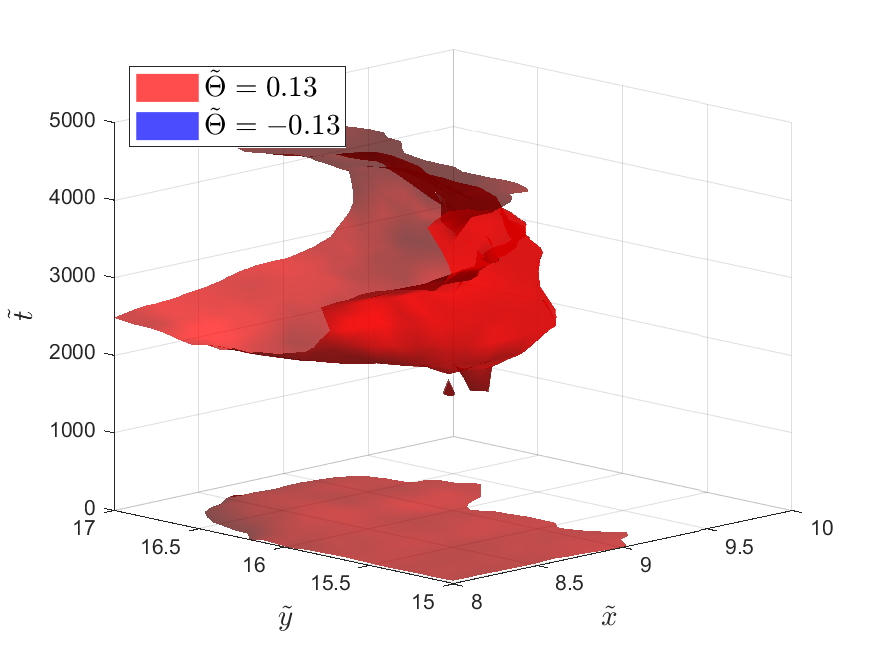}\vspace{0.1cm}
\includegraphics[width=0.48\textwidth]{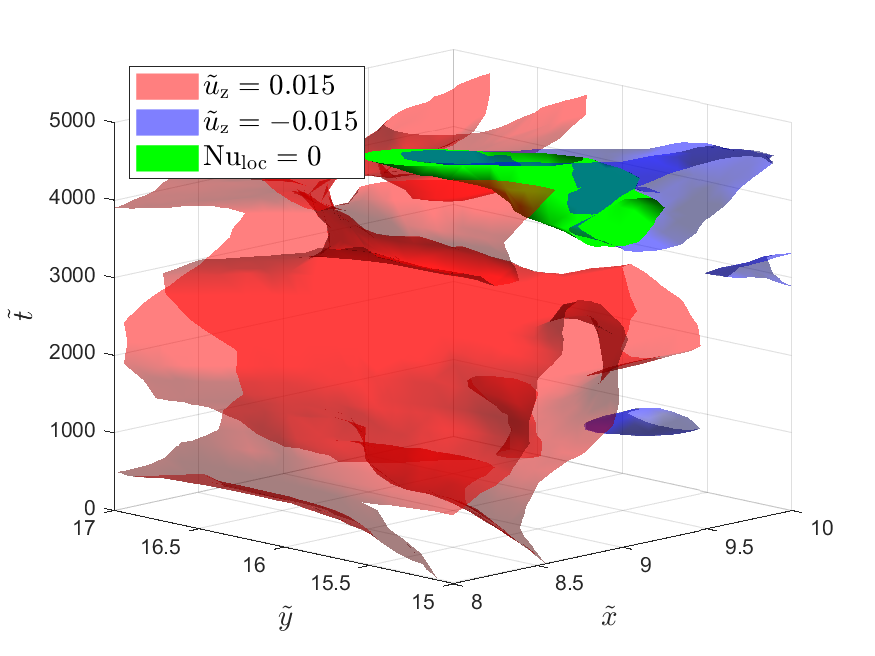}
\caption{Isosurfaces of $\tilde{\Theta}$ (left) as well as of $u_{\rm{z}}$ and $\rm{Nu}_{\rm{loc}}$ (right) in sections covering $\Delta \tilde{x} \times \Delta \tilde{y}= 4 \times 4$ (top) and $\Delta \tilde{x} \times \Delta \tilde{y}= 2 \times 2$ (bottom) for the experimental data obtained from the measurement at $\mathrm{Ra}=2 \times 10^5$.}
\label{fig:Isosurfaces_zoom}
\end{figure}

A more detailed look into one event with heat transfer towards the bottom plate can be seen in figure~\ref{fig:Isosurfaces_zoom}. In the top left panel a region with the temperature below the mean temperature is displayed. It is visible for $\tilde{t} \in [1000, 3500]$. If at the same time the vertical velocity points downwards, the local Nusselt number is positive. However, at the right hand side of the figure it can be seen in the isosurfaces of the vertical velocity component, that at a certain time instant the velocity points upwards and thus colder fluid will be transported towards the cooling plate. In this case, the local Nusselt number becomes negative, indicated by the green isosurface of the local Nusselt number for $\tilde{t} \in [2500, 3500]$. Later the velocity still points upwards but the structures in the temperature field have reorganized and the local temperature is not anymore colder than the mean temperature, such that the Nusselt number becomes positive again. From the current data it seems that this is a process that can be seen more often in the vicinity of the upwelling ridges. To summarize, a gradual evolution of the turbulent superstructures is clearly observable as our space-time analysis suggests. 

\section{Conclusion and outlook}
\label{sec:conclusion}
In this work, the structure and evolution of turbulent superstructures in a horizontally extended Rayleigh-B\'{e}nard convection layer have been investigated in a controlled laboratory experiment. Particle image velocimetry and thermometry were combined for the measurements in a water-filled Rayleigh-B\'{e}nard cell with dimensions of $l \times w \times h = 700\,\mathrm{mm} \times 700\,\mathrm{mm} \times 28\,\mathrm{mm}$. The aspect ratio is $\Gamma=l/h=25$, the Prandtl number ${\rm Pr}=7$, and the range of Rayleigh numbers $2\times 10^5\le  {\rm Ra}\le 7\times 10^5$; thus the experimental results are directly comparable with existing results of direct numerical simulations. Here, we accessed the field structure of the three velocity components and the temperature in their joint dynamical evolution in a wider observation area of about $451\,\mathrm{mm} \times 468\,\mathrm{mm}$ centered in the bulk of the cell to study the spatio-temporal organization of the turbulent heat transfer. Due to its transparent sidewalls and the transparent cooling plate made of glass the experimental setup provides full optical access to the turbulent flow. Simultaneous measurements of the temperature and velocity field in a large section of the horizontal midplane of the cell are thus feasible using thermochromic liquid crystals as tracer particles. Based on their temperature-dependent color shade the temperature field has been determined, while their temporal displacement has been evaluated to calculate both the horizontal and the vertical components of the velocity field in the observation plane. 

The time-averaged fields of the temperature and the vertical velocity component, obtained from the present experiments and the DNS, clearly display turbulent superstructures of convection and agree qualitatively well. However, quantitative differences between the experimental and numerical results, in particular with respect to the characteristic wavelength of the large-scale patterns and the magnitude of the global Nusselt number, have been found. While the characteristic length scales of the turbulent superstructures are larger in the experiment, the values of the Nusselt number ${\rm Nu }$ are smaller compared to the results of the DNS. Both can be traced back to the deviations from the ideal isothermal boundary conditions at the top plate of the convection cell, which we discussed in detail in the text by an estimation of the corresponding Biot numbers. Despite the slight differences of the magnitude of the Nusselt number, the power law exponent of the transport law ${\rm Nu}({\rm Ra})$ is in the same range as those of the DNS and other studies of thermal convection in water \citep{Ahlers2009,Chilla2012}. 

The joint measurement of the temperature and of the vertical velocity component has been applied to study the local convective heat flux in the experiment, thereby clearly demonstrating that the skeleton of turbulent superstructures contributes significantly to the turbulent heat transfer from the bottom to the top of the cell. The contribution of the turbulent superstructures to the heat flux is therefore quantified by means of the local convective Nusselt number, which reaches its maxima within the localized up-- and downdrafts that compose to the superstructure patterns in the flow. 

Since the measurements of the temperature and velocity fields cover up to approximately $10^4$ convective free-fall times, our investigations also allow to study the reorganization of the superstructures for the different Rayleigh numbers. Their re-arrangement is analyzed here by means of the correlation coefficient of the time-averaged temperature fields. It is seen that the superstructures in the experiment have a preferential arrangement, most probably due to the slightly inhomogeneous temperature at the horizontal top plate. Nevertheless, the superstructures exhibit a clear gradual reorganization, which is in good agreement with the findings obtained in the previous numerical studies by \citet{Pandey2018} and \citet{Fonda2019}.

Despite the non-ideal boundary conditions at the top plate, the presented results prove the concept of an experimental study of the long-term evolution of turbulent superstructures in Rayleigh-B\'{e}nard convection. A further improvement of the thermal boundary conditions at the top plate of the experimental facility is one point of our future work. An alternative approach to the determination of the structures, which are connected to the local convective heat flux, would be by three-dimensional, three-component volumetric (3D3C) measurements in a small volume. Again they would have to be combined with TLC measurements and require a larger number of simultaneously operating cameras. These studies are currently underway and will be reported elsewhere.  

\section*{Acknowledgements}
Funding of this work by the Deutsche Forschungsgemeinschaft (DFG) within the Priority Programme DFG-SPP 1881 ``Turbulent Superstructures'' and by the Carl Zeiss Foundation within the project no. P2018-02-001 ``Deep Turb--Deep Learning in and of Turbulence'' is acknowledged. Furthermore, the authors wish to thank Alexander Thieme and Jens Fokken for the technical support in the laboratory and Philipp Vieweg for discussions and the additional analysis of his simulation data.

\section*{Declaration of Interests} The authors report no conflict of interest.


\end{document}